\documentclass[preprint,preprintnumbers,amsmath,amssymb,floatfix,nofootinbib]{revtex4}
\usepackage{epsf}
\usepackage{graphicx}
\begin{document}          

\vskip 2in
\hskip 5.2in \hbox{\bf YITP-SB-10-04}

\title{\vspace*{0.7in}
 Weak Corrections to Associated Higgs-Bottom Quark Production}

\author{S.~Dawson$^{a}$ and P.~Jaiswal$^{a,b}$}

\affiliation{
$^a$Department of Physics, Brookhaven National Laboratory, 
Upton, NY 11973, USA \\
$^{b}$Yang Institute for Theoretical Physics, Stony Brook University, Stony Brook, NY 11790, USA
\vspace*{.5in}}

\begin{abstract}

In models with an enhanced coupling of the Higgs boson to the bottom quark, the 
dominant production mechanism in hadronic collisions 
is often the partonic sub-process,
$bg\rightarrow bH$.
 We derive the weak
corrections to this process and show that they can be
accurately approximated by an ``Improved Born Approximation''.
At the Tevatron, these corrections are negligible
and are dwarfed by PDF and scale uncertainties for $M_H < 200~GeV$.
At the LHC,  
the weak corrections are small for $M_H < 500~GeV$.
For
  large Higgs boson masses, the corrections become
significant and
are $\sim ~18\%$ for $M_H\sim 1~TeV$ at $\sqrt{s}=10~TeV$.

\end{abstract}

\maketitle
\newpage

\section{Introduction}

The search for the Higgs boson is 
one of the most important tasks for both the Fermilab Tevatron and the CERN
Large Hadron Collider. The Standard Model requires
a single scalar Higgs boson, with well defined properties except for
its mass.  In this case, the Higgs boson will be discovered at either
the Tevatron or the LHC, with the discovery channel depending strongly
on the Higgs boson mass.  
In the Standard Model, the production of a Higgs
boson in association with $b$ quarks is never a discovery channel
due to the small $b$ quark- Higgs boson 
Yukawa coupling.  However, in non-standard
models of electroweak symmetry breaking
with a light Higgs boson,
 the coupling of the Higgs boson to the $b$ quark is
often enhanced and
the channels $b{\overline b}\rightarrow H$ and
 $gb\rightarrow bH$ become  important  
modes\cite{Dawson:2005vi,Dawson:2004sh,Campbell:2004pu,Dittmaier:2003ej,Dawson:2003kb,Campbell:2002zm,Maltoni:2005wd,Dicus:1998hs,Maltoni:2003pn}.
A  familiar example of such a model is the MSSM with
large $\tan\beta$ where
enhancements by orders of magnitude over the
Standard Model prediction are possible in some parameter
regions\cite{Brein:2003df,Field:2003yy,Carena:1998gk,Carena:2007aq}.

The hadronic production rate for the associated
 production of a Higgs boson and a $b$ quark is well 
understood\cite{Dawson:2004sh,Dawson:2005vi,Campbell:2004pu,Dittmaier:2003ej,Dawson:2003kb,Maltoni:2005wd,Dicus:1998hs,Campbell:2002zm,Maltoni:2003pn}. 
The calculation can be done in either a $4$- flavor or a $5$- flavor number
parton distribution function (PDF)
 scheme, which represent different orderings of perturbation theory.
In the $4$- flavor number PDF scheme, the lowest order processes 
for producing a Higgs boson and a $b$ quark are $gg\rightarrow
 b {\overline b}H$ and 
$ q {\overline q}\rightarrow b {\overline b} H$.
Alternatively, in the  $5$- flavor number scheme, the $b$ quark is 
treated as
 a parton
and large logarithms of the form $\ln({M_H\over m_b})$
are absorbed into $b$ quark parton distribution 
functions\cite{Barnett:1987jw,Olness:1987ep}. 
In this scheme, the lowest order process for producing
a Higgs boson in association with $b$ quarks is $b{\overline b}
\rightarrow H$ when no $b$ quarks are tagged in the final state, and 
 $b g\rightarrow b H$ when a single outgoing $b$ quark 
is tagged.  Within the uncertainties, the $4-$ and $5-$ flavor
number schemes give equivalent results for the NLO QCD corrected
rate for associated $b$ quark-Higgs production\cite{Campbell:2004pu}.

We work in the $5-$ flavor number scheme for simplicity and consider
the associated production of a $b$ quark and a Higgs boson.
The $b {\overline b}\rightarrow
H$ rate is known to NNLO QCD\cite{Harlander:2003ai}, along with the full
electroweak and SUSY QCD corrections\cite{Dittmaier:2006cz,Hollik:2006vn}.  
When an outgoing $b$ quark is tagged, the rate is lower, but
the background is significantly reduced, making
this an important channel. 
Both the NLO QCD\cite{Dawson:2005vi,Dawson:2003kb,Campbell:2004pu,Dawson:2004sh,Dittmaier:2003ej,Campbell:2002zm} and the
SUSY QCD (SQCD) corrections from gluino-squark 
loops in the case of the 
MSSM\cite{Dawson:2007ur}  are known for $bH$ production.
Furthermore,  the Tevatron experiments have produced
limits on $bH$ production in the 
MSSM\cite{Abazov:2008hh,Abazov:2008zz} which can be
interpreted in terms of the fundamental properties of the model.  

In this paper, we compute the Standard Model weak 
corrections to the $bg\rightarrow b H$
process and compare them with the scale and PDF uncertainties of the
NLO QCD corrected rates.  
We also compare our results with an approximation
where the dominant corrections arise from the
 on-shell corrections to the
${\overline b} b H$ vertex (Improved Born Approximation). 
Section II contains the theoretical framework for the weak corrections. 
We retain  the effects of a non-zero $b$ quark mass everywhere.  
Numerical results are given in Section III and 
conclusions in Section IV.

\section{Theoretical Framework}
\label{sec:theory}
In this paper, we consider the Standard Model process of associated $b$
quark- Higgs boson
production.  Our results can be generalized in a straightforward manner
to models with non-standard $b$ quark - Higgs boson
couplings.
The tree level coupling of a $b$ quark to a Standard Model Higgs 
boson, $H$,  is given
by 
\begin{equation}
L_{YUK}=-g_{b0}{\overline b}_0 b_0 H_0\, ,
\label{yukdef}
\end{equation} 
where the subscript, `0', denotes the unrenormalized quantity and
\begin{equation}
g_{b_0}=
{m_{b0}\over 2 M_{W_0}}
{e_0\over s_{W_0}}\, .
\label{lren}
\end{equation}
We work in an on-shell scheme where the weak mixing angle is a
derived quantity and is defined in terms of the physical gauge
boson masses,
\begin{equation}
\sin^2\theta_W\equiv s_W^2=1-{M_W^2\over M_Z^2}\, .
\label{sindef}
\end{equation}

\begin{figure}[t]
\begin{center}
\includegraphics[scale=0.8]{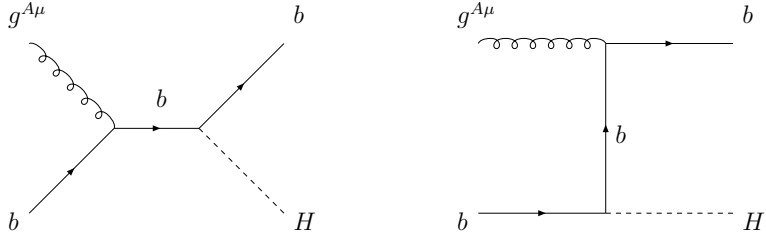}
\caption[]{
Lowest order Feynman diagrams for the process $bg\rightarrow bH$.}
\label{fg:lo}
\end{center}
\end{figure}

The lowest order Feynman diagrams
for the process $b(p_1)+ g(q_1)\rightarrow b(q_2)+ H(p_2)$  
are shown in Fig.~\ref{fg:lo}.  The
resulting Born cross section is\cite{Campbell:2002zm},
\begin{eqnarray}
{d\sigma (bg\rightarrow b H)(\mu_R)_0\over dt}
&=&{1\over (s-m_b^2)^2} {\alpha_s(\mu_R)\over 24} 
{\overline {g_b}(\mu_R)}^2 \biggl\{-{M_H^4+u^2\over s_1 t_1}
+{2 m_b^2\over s_1^2 t_1^2}\biggl[ 4 u t_1s_1+M_H^2(M_H^2-u)^2\biggr]
\nonumber \\
&&
-8{m_b^4\over s_1^2 t_1^2}\biggl(M_H^2-u\biggr)^2
\biggr\}
\, ,
\label{eq:lo}
\end{eqnarray}
where $s_1=(p_1+q_1)^2-m_b^2=s-m_b^2$, ~$t_1=(p_1-p_2)^2-m_b^2,$ and $u=(p_1-q_2)^2$,  
are the usual Mandelstam variables and the scale $\mu_R$ is the arbitrary
renormalization scale.  (The cross section for the charge conjugate
process, ${\overline b} g\rightarrow {\overline b} H$, is identical to
Eq. \ref{eq:lo}.)  In the limit $m_b\rightarrow 0$, the tree level
contribution to $bg\rightarrow bH$ vanishes and the first non-zero contributions
are a subset of the $1-$ loop amplitudes computed in this work.  The $m_b\rightarrow 0$
limit has been considered in Refs. \cite{Mrenna:1995cf,Boudjema:2008zn} and we will comment on
 the numerical effects of
this limit in Section III.  In this paper, however, we keep $m_b$ nonzero everywhere.

In Eq. \ref{eq:lo}, the Yukawa coupling, 
${\overline{g_b}(\mu_R)}$, is expressed in terms of the
$1$-loop renormalization group improved running 
${\overline{MS}}$ mass for the $b$ quark, ${\overline{m_b}(\mu_R)}$. 
For the decay $H\rightarrow
b {\overline b}$, the ${\cal O}(\alpha_s \log({m_b\over M_H}
))$ contributions can be 
absorbed into ${\overline{m_b}(M_H)}$\cite{Kniehl:1994ju,Djouadi:2005gi},
motivating our use of the running mass.  The
${\cal{O}}(\alpha_S)$ NLO predictions for the production process,
$bg\rightarrow bH$, however, depend sensitively on this 
choice for $g_b$\cite{Dawson:2005vi}.

\subsection{Renormalization}
As input parameters in
the electroweak sector,
 we take $\alpha(0)$, $M_Z$, and  $G_\mu$, along with 
the Higgs boson
and fermion masses.  The $W$ mass is then a derived quantity.  At tree
level,
\begin{equation}
M_{W0}^2={\pi\alpha_0\over\sqrt{2} G_{\mu0}(1-M_{W0}^2/M_{Z0}^2)}\, .
\label{mwdef}
\end{equation}
The gauge boson $2$ -point functions are defined as,
\begin{equation} 
\Pi_{XY}^{\mu\nu}(p^2)=
g^{\mu\nu}\Pi_{XY}(p^2)+p^\mu p^\nu B_{XY}(p^2)\, ,
\end{equation}
where $XY=WW,ZZ,\gamma \gamma$
and $\gamma Z$.  Analytic results for the 
Standard Model 
gauge boson and Higgs contributions
can  be found in Refs. \cite{Hollik:1988ii,Bardin:1999ak}\footnote{A convenient compilation
of the gauge boson and Higgs contributions to the gauge boson $2$-point
functions employing our conventions
can be found in the appendix of Ref. \cite{Chen:2008jg}.}
 and for the fermion contributions in the appendix of Ref. \cite{Chen:2003fm}.
The gauge boson mass renormalizations ($M_V=M_W,M_Z)$
 are  defined by the on-shell condition,
\begin{eqnarray}
M_{V_0}^2&=&M_V^2\biggl(1+{\delta M_V^2\over M_V^2}\biggr)= M_V^2\biggl(1+
{\Pi_{VV}(M_V^2)\over M_V^2}
\biggr)\, .
\end{eqnarray}

The electromagnetic charge renormalization is determined
 from Thompson scattering by\footnote{This relation effectively defines our convention for the sign
of $\Pi_{\gamma Z}$.},
\begin{eqnarray}
\alpha_0&=&\alpha\biggl(1+{\delta \alpha\over \alpha}\biggr)\nonumber \\
{\delta\alpha\over \alpha}&=&\Pi^\prime_{\gamma\gamma}(0)+2{s_W
\over c_W}{\Pi_{\gamma Z}(0)\over M_Z^2}\, ,
\end{eqnarray}
where the contribution from gauge bosons, leptons, and the top quark  is
given by $\Pi^\prime_{\gamma \gamma}(0)=\partial \Pi_{\gamma\gamma}(p^2)/\partial p^2 \mid_{p^2=0}$.  The light fermions contribute large logarithms to
$\Pi^\prime_{\gamma \gamma}(0)$ which we estimate by 
$\Pi_{\gamma\gamma}^\prime(0)\mid_{had}=\Pi_{\gamma\gamma}(M_Z^2)/M_Z^2
+\Delta\alpha^5\mid_{had}$, with 
$\Delta\alpha^5\mid_{had}=.02761\pm .0036$\cite{Burkhardt:2001xp}.

The Fermi constant $G_\mu$ is found from muon decay and,
 by definition, QED corrections are
included in the measured value\cite{Sirlin:1981yz,Sirlin:1980nh},
\begin{equation}
G_{\mu0}=G_\mu\biggl(1+{\delta G_\mu\over G_\mu}\biggr)\, ,
\end{equation}
where,
\begin{eqnarray}
{\delta G_\mu\over G_\mu}&=&-{\Pi_{WW}(0)\over M_W^2}+\delta_{V-B}
\nonumber \\
\delta_{V-B}&=& -{\alpha\over 4\pi s_W^2}\biggl(
6+{7-4 s_W^2\over 2 s_W^2}\log(c_W^2)\biggr)+{2\over s_Wc_W}{\Pi_{\gamma Z}(0)\over M_Z^2}\, .
\label{gren}
\end{eqnarray}
In the second line of Eq. \ref{gren}, $\delta_{V-B}$ is the contribution
from vertex and box diagrams.  For consistency, the $W$ mass in this equation should
be found from the tree level prediction, Eq. \ref{mwdef}.

The weak mixing angle renormalization is defined,
\begin{equation}
s_{W0}^2=s_W^2\biggl(1+{\delta s_W^2\over s_W^2}\biggr)\, .
\end{equation}
Using the on-shell definition of  Eq. \ref{sindef},
\begin{equation}
{\delta s_W^2\over s_W^2}
={c_W^2\over s_W^2}
\biggl(
-{\Pi_{WW}(M_W^2)\over M_W^2}+{\Pi_{ZZ}(M_Z^2)\over M_Z^2}
\biggr)\, .
\end{equation}

The fermion renormalization proceeds in the usual manner\footnote{$b_{L,R}\equiv
(1\mp \gamma_5) b$ and $P_{L,R}\equiv (1\mp \gamma_5)/2$.},
\begin{eqnarray}
m_{b0}&=& m_b+\delta m_b\nonumber \\
b_{L,R}^0 &\equiv & \sqrt{Z_{L,R}} b_{L,R} =
\biggl(1+{\delta Z_{L,R}\over 2}\biggr)b_{L,R}
\, .
\label{brenorm}
\end{eqnarray}
The one-loop $b$ quark self- energy is,
\begin{eqnarray}
\Sigma_b(k^2)
& \equiv & k \biggl[P_L\biggl(\Sigma_V^L(k^2)+Re(\delta Z_L)\biggr)
+P_R\biggl(\Sigma_V^R(k^2)+Re(\delta Z_R)\biggr)\biggr]\nonumber \\
&& -m_b\biggl[ {\delta m_b\over m_b}-\Sigma_S(k^2)+{\delta Z_L+\delta Z_R^\dagger\over 2}P_L
+{\delta Z_R+\delta Z_L^\dagger\over 2}P_R\biggr]\, ,
\end{eqnarray}
Imposing the on-shell conditions,
\begin{eqnarray}
{\delta m_b\over m_b}&=&Re\biggl\{\Sigma_S(m_b^2)+{\Sigma_V^L(m_b^2)+\Sigma_V^R(m_b^2)\over 2}
\biggr\}
\nonumber \\
\delta Z_L&=&-Re\biggl\{\Sigma_V^L(m_b^2)
-m_b^2\biggl(\Sigma_V^{L}(m_b^2)^\prime+\Sigma_V^R(m_b^2)^\prime
+2\Sigma_S(m_b^2)^\prime\biggr\}\nonumber \\
\delta Z_R&=&-Re\biggl\{\Sigma_V^R(m_b^2)
-m_b^2\biggl(\Sigma_V^{L}(m_b^2)^\prime+\Sigma_V^R(m_b^2)^\prime
+2\Sigma_S(m_b^2)^\prime\biggr\}\, ,
\end{eqnarray}
where $\Sigma(m^2)^\prime \equiv \partial \Sigma(p^2)/\partial p^2 \mid_{p^2=m^2}$.
Analytic results for the expressions in Eq. \ref{brenorm} are
given in Refs. \cite{Hollik:1988ii,Bardin:1999ak}. 

The Yukawa coupling renormalization is defined as,
\begin{equation}
g_{b_0}=
 g_b\biggl(1+{\delta g_b\over g_b}\biggr)\, .
\end{equation}
Combining the above results, 
the one-loop electroweak counterterm corresponding to Eq. \ref{yukdef}
 is then,
\begin{eqnarray}
L_{YUK}&=&-g_b\biggl(1+\delta_{CT}\biggr){\overline b} b H
\nonumber \\
\delta_{CT}&=&{\delta g_b\over g_b}+{\delta Z_H\over 2}+{\delta Z_L+
\delta Z_R\over 2}
\label{count}
\end{eqnarray}
where,
\begin{eqnarray}
{\delta g_b\over g_b}&=&
{\delta m_b\over m_b}+{1\over 2}{\delta G_\mu\over G_\mu}\nonumber \\
&=&{\delta m_b\over m_b}-{\delta M_W\over M_W}
+{\delta e\over e}-{\delta s_W\over s_W}\, .
\label{dgb}
\end{eqnarray}

Finally, we need the $1$-loop corrected value for the $W$ 
mass\cite{Marciano:1983wwa},
\begin{eqnarray}
M_W^2&=&{\pi\alpha\over \sqrt{2}G_\mu s_W^2}{1\over 1- \Delta r}
\nonumber \\
&=&
{M_Z^2\over 2}\biggl(1+\sqrt{1-{4\alpha\over \sqrt{2}G_\mu M_Z^2}
(1+\Delta r)}\biggr)\, ,
\label{wdef}
\end{eqnarray}
where,
\begin{eqnarray}
\Delta r&=&\Pi_{\gamma\gamma}^\prime (0)
-2{c_W\over s_W}{\Pi_{\gamma Z}(0)\over M_Z^2}
+{\Pi_{WW}(0)-\Pi_{WW}( M_W^2)\over M_W^2}
\nonumber \\
&&-{c_W^2\over s_W^2}\biggl({\Pi_{ZZ}(M_Z^2)\over M_Z^2}-
{\Pi_{W}(M_W^2)\over M_W^2}\biggr)
 +{\alpha\over 4\pi s_W^2}\biggl(
6+{7-4 s_W^2\over 2 s_W^2}\log(c_W^2)\biggr)
\, .
\end{eqnarray}

\subsection{Electroweak corrections to $H\rightarrow b {\overline b}$}

\begin{figure}[t]
\begin{center}
\includegraphics[scale=1.0]{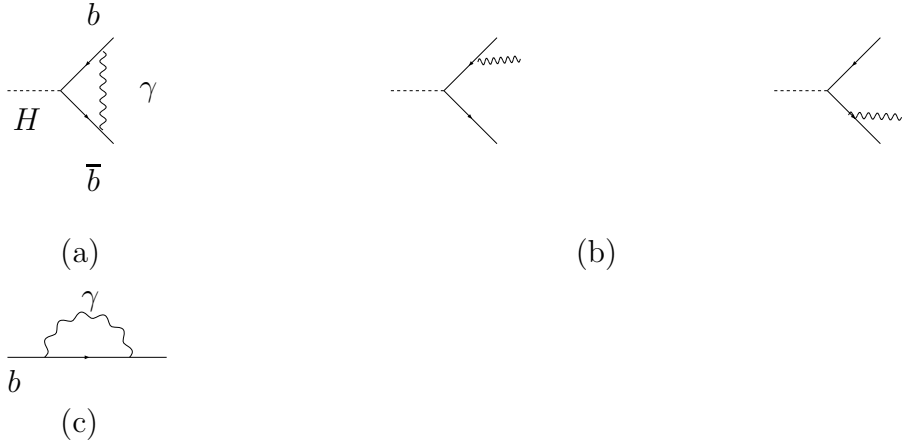}
\caption[]{(a) Virtual QED corrections
to the decay $H\rightarrow b {\overline b}$, (b)
Real photon emission contributions to $H\rightarrow b {\overline b}\gamma$,
and (c) Feynman diagram contributing to the QED counterterms for $H\rightarrow
b {\overline b}$.}
\label{fg:hbb_feyn}
\end{center}
\end{figure}

\begin{figure}[t]
\begin{center}
\includegraphics[scale=0.6]{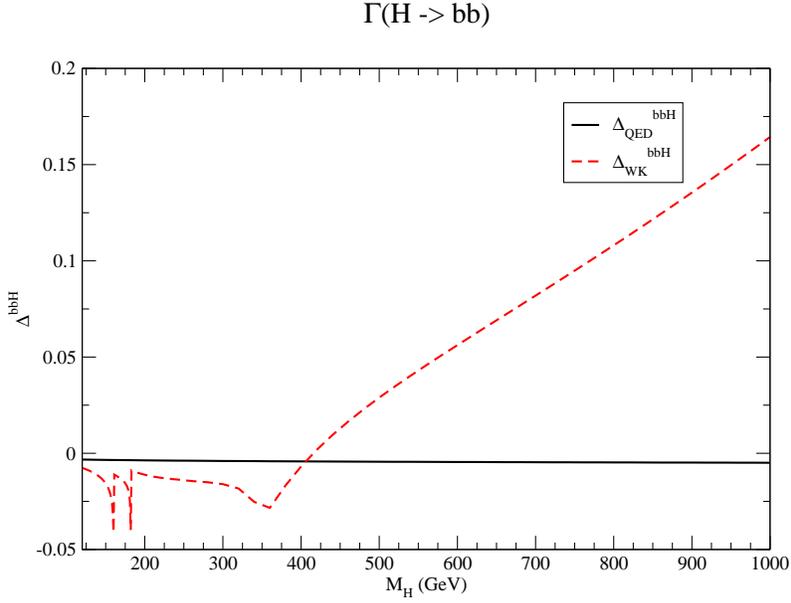}
\caption[]{Pure QED (black solid)  and weak (red dashed)  corrections to 
the decay $H\rightarrow b {\overline b}$ as defined in Eq. \ref{hdecdef}.}
\label{fg:hbb}
\end{center}
\end{figure}

The calculation of the
electroweak corrections to $gb\rightarrow bH$ has many common features
with that for $H\rightarrow b {\overline b}$ and so we review the
decay process briefly.  This discussion will also serve to make clear our
separation of the  QED and weak contributions.
The electroweak corrections contain both pure QED photonic contributions
and the remaining weak corrections and the two contributions are separately gauge
invariant\cite{Kniehl:1991ze}.  

The corrections to $H\rightarrow b {\overline b}$ can be parameterized as,
\begin{equation}
\Gamma(H\rightarrow b {\overline b})
=\Gamma(H\rightarrow b{\overline b})_0\biggl(1
+\Delta_{QCD}^{bbH}+\Delta_{QED}^{bbH}
+\Delta_{WK}^{bbH}\biggr)\, ,
\label{hdecdef}
\end{equation}
where $\Gamma(H\rightarrow b{\overline b})_0$ is the tree level result, but
evaluated with a running $b$ quark mass as described above. 
The QCD corrections are known to ${\cal O}(\alpha_s^2)$, including all
top quark mass effects\cite{Larin:1995sq,Chetyrkin:1995pd}.

The QED corrections are found from the virtual photon diagram of Fig \ref{fg:hbb_feyn}a, 
the real photon emission
diagrams of Fig. \ref{fg:hbb_feyn}b, and the counterterms derived from Fig. \ref{fg:hbb_feyn}c,
\begin{equation}
\Delta_{QED}^{bbH}=
\Delta_{Virt,QED}^{bbH}
+\Delta_{CT,QED}^{bbH}
+\Delta_{Real,QED}^{bbH}\, .
\end{equation}
  The virtual photon
contribution in $N=4-2\epsilon$ dimensions is,
\begin{eqnarray}
\Delta_{Virt,QED}^{bbH}&=&  {\alpha\over \pi} Q_b^2
\biggl({4\pi\mu^2\over m_b^2}\biggr)^\epsilon\biggl\{
{2\over\epsilon}+
{1+\beta^2\over \beta}\biggl[
-{1\over 2 \epsilon}L
-{L^2\over 4}+{\pi^2\over 3}+
\nonumber \\ &&
Li_2\biggl({1-\beta\over 1+\beta}\biggr)
-L\ln\biggl({1+\beta\over 2\beta}\biggr)\biggr]
-\biggl({1-\beta^2\over \beta}\biggr)L 
+3\biggr\}\, ,
\end{eqnarray}
where $\beta=\sqrt{1-4m_b^2/M_H^2}$ and $L=\log((1-\beta)/(1+\beta))$.
This is in agreement with Refs. \cite{Braaten:1980yq,Drees:1990dq} 
with the replacement ${4\over 3} \alpha_s 
\rightarrow Q_b^2 \alpha$ and $Q_b=-{1\over 3}$. 

To find the counterterms, we need the photonic contributions to Eqs. \ref{count} and \ref{dgb}.
By definition, $\delta G_\mu$ contains only the weak contributions.
Similarly, the Higgs boson self-energy does not receive contributions
from diagrams with photons in the internal loops.  Thus,
we have to separate the QED contributions to 
$\delta m_b$, $\delta Z_L$ and
$\delta Z_R$ which come only 
 from Fig. \ref{fg:hbb_feyn}c .  The results are well known,
\begin{eqnarray}
\delta Z_{L,QED}=\delta Z_{R,QED}
&=& -{\alpha\over 4 \pi} Q_b^2
\biggl({4\pi \mu^2\over m_b^2}\biggr)
^\epsilon \Gamma(1+\epsilon)\biggl({3\over \epsilon}+4\biggr)
\nonumber \\
\biggl({\delta m_b\over m_b}\biggr)_{QED}&=&
 -{\alpha\over 4 \pi} Q_b^2
({4\pi \mu^2\over m_b^2}\biggr)
^\epsilon \Gamma(1+\epsilon)\biggl({3\over \epsilon}+4\biggr)
\, .
\end{eqnarray}
The QED counterterm is then,
\begin{eqnarray}
\Delta_{CT,QED}&=& 2\biggl\{ {\delta m_b\over m_b}+{\delta Z_L+\delta Z_R\over 2}\biggr\}
\nonumber \\
&=&-{\alpha\over \pi}Q_b^2
\biggl({4\pi \mu^2\over m_b^2}\biggr)^\epsilon 
\Gamma(1+\epsilon)
\biggl\{{3\over \epsilon}
+4\biggr\}\, .
\end{eqnarray}

Finally, we need the real photon emission 
contributions from Fig \ref{fg:hbb_feyn}b,
\begin{equation}
\Delta_{Real,QED}=
{\alpha\over \pi}Q_b^2
\biggl({4\pi \mu^2\over m_b^2}\biggr)^\epsilon 
\Gamma(1+\epsilon)\biggl\{{1\over\epsilon}\biggl[1+{1+\beta^2\over 2 \beta}L
\biggr]+F(M_H,\beta)\, ,
\end{equation}
where $F(M_H.\beta)$ is finite and an analytic form can be found 
in Refs.\cite{Drees:1990dq,Braaten:1980yq,Kniehl:1991ze,Dabelstein:1991ky}.

 The QED contributions enumerated above are recognized by the explicit over-all factors
of $Q_b^2$.  The remaining weak corrections to $H\rightarrow b {\overline b}$,
$\Delta_{WK}^{bbH}$, are given analytically
in Refs. \cite{Kniehl:1991ze,Dabelstein:1991ky} 
and are found from diagrams with $W$'s,~$Z$'s, 
and Goldstone bosons, along
with top quark contributions. 
In Fig. \ref{fg:hbb} we plot the QED and weak contributions to $\Gamma(H
\rightarrow b {\overline b})$.  
The QED corrections are always ${\cal O}(10^{-3})$
and can safely be neglected for all practical purposes\footnote{The spikes 
at the $W^+W^-$ and $ZZ$ thresholds are softened if the complex mass
scheme is employed\cite{Denner:2005fg,Passarino:2010qk}.}.

\subsection{One-Loop Corrections}

\begin{figure}[t,b]
\begin{center}
\includegraphics[scale=1.0]{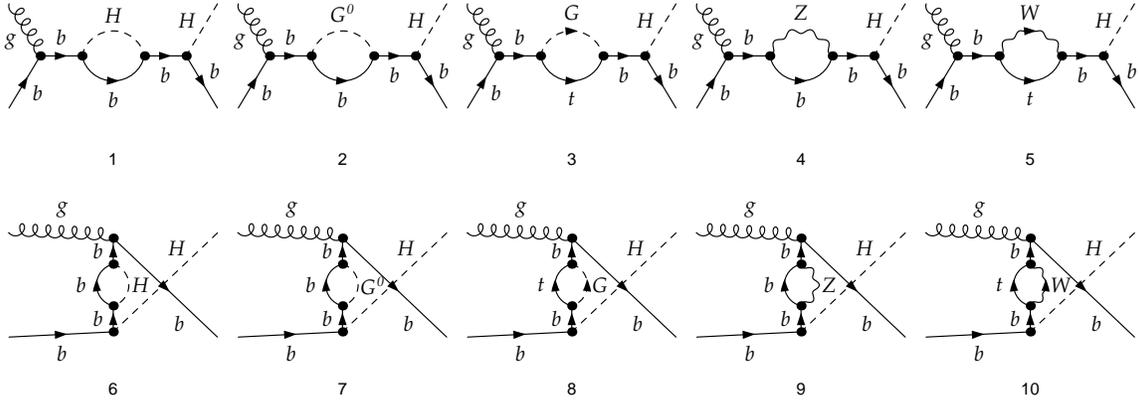}
\caption[]{
Self energy diagrams contributing to the weak corrections
to $b g\rightarrow bH$. }
\label{fg:self}
\end{center}
\end{figure}

\begin{figure}[t]
\begin{center}
\includegraphics[scale=1.0]{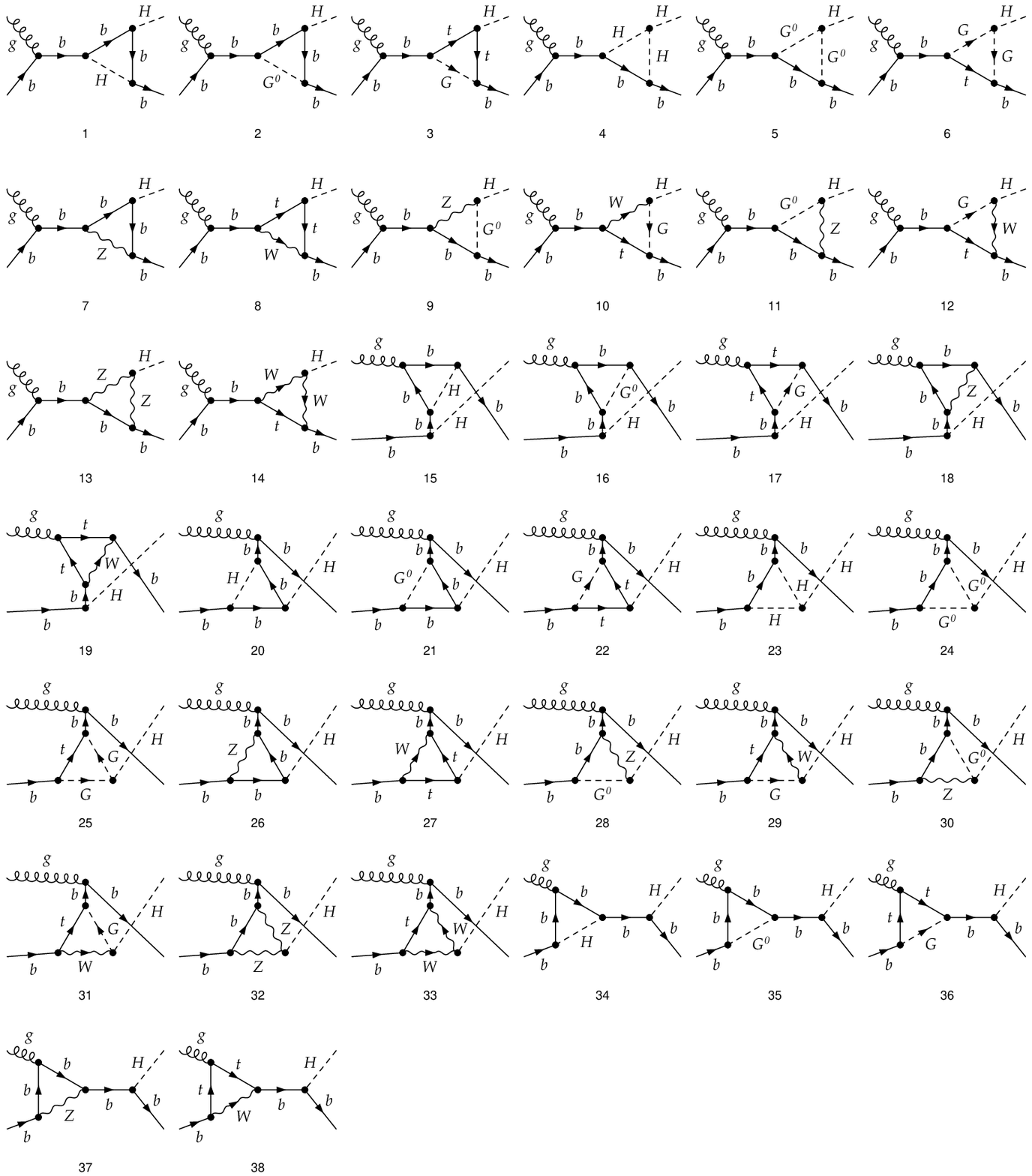}
\caption[]{
Vertex diagrams contributing to the weak corrections
to $b g\rightarrow bH$. }
\label{fg:vert}
\end{center}
\end{figure}

\begin{figure}[t]
\begin{center}
\includegraphics[scale=1.0]{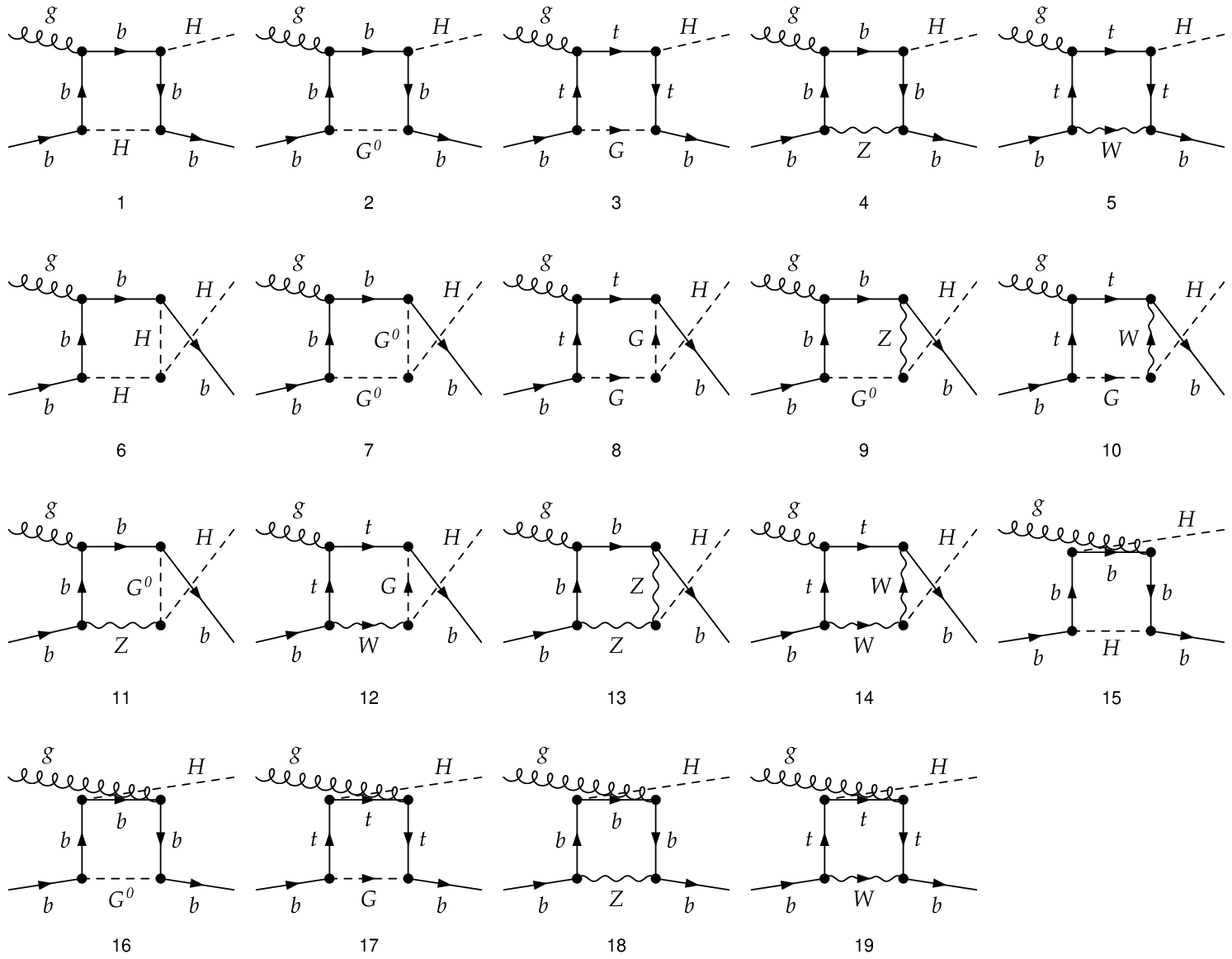}
\caption[]{
Box diagrams
contributing to the weak corrections
to $b g\rightarrow bH$.}
\label{fg:box}
\end{center}
\end{figure}

The one loop weak corrections to the process $bg\rightarrow
b H$ consist of self energy, 
vertex, and box diagrams (Figs. \ref{fg:self}-\ref{fg:box}),
along with the counterterms given explicitly 
in Eq. \ref{count}. Our results can be expressed as,
\begin{equation}
\sigma(bg\rightarrow bH)_{NLO}=\sigma(bg\rightarrow bH)_0\biggl(
1+\Delta_{QCD}+\Delta_{QED}+\Delta_{WK}\biggr)\, ,
\label{cordef}
\end{equation}
where $\sigma_0$ is the Born cross section of Eq. \ref{eq:lo}
 evaluated with the
$1-$ loop renormalization group improved value for ${\overline{g_b}}$.
 
The purely photonic QED corrections consist of vertex and box contributions
from internal photons along with the corresponding counterterms, 
real radiation from $bg\rightarrow bH\gamma$, and the process
involving photons in the initial state, $\gamma b\rightarrow bgH$.
The ${\cal O}(\alpha)$ QED corrections to
 $bg\rightarrow bH$  can be found from
the corresponding QCD 
corrections by making
the substitution ${4\over 3}\alpha_s\rightarrow 
\alpha Q_b^2$\cite{Dawson:2005vi,Dawson:2003kb,Campbell:2004pu,Dawson:2004sh,Dittmaier:2003ej,Campbell:2002zm}.However,  
evaluating the process  $\gamma b\rightarrow bgH$
requires the use of a PDF set which
includes initial state photons.\footnote{The most modern set of PDFs
which include initial state photons are 
the MRST2004qed PDFs\cite{Martin:2004dh}.}
This contribution is  expected to be quite
small since potentially large logarithms from 
initial state
collinear photon emission 
are absorbed into the PDFs. 
We further note that the QED contributions
to the $b{\overline b}\rightarrow H$ process \cite{Dittmaier:2006cz},
and to the
corresponding decay $H\rightarrow 
b {\overline b}$\cite{Kniehl:1991ze,Dabelstein:1991ky} discussed
above,
are known to be 
less than $1~\%$.  As this is considerably smaller than the PDF
and scale uncertainties which we present in the next section,
we do not provide numerical results for
the pure QED corrections to $bg\rightarrow bH$, but evaluate only
the weak corrections.

The
Feynman diagrams are generated using FeynArts\cite{Hahn:2000kx} and
the interference with the tree level amplitude is
evaluated numerically in Feynman gauge
using FormCalc and LoopTools\cite{Hahn:1998yk}.
We retain  a non-zero bottom quark mass 
everywhere.

\subsection{Large Higgs Mass Limit}

The contributions to the weak corrections in the large Higgs mass
limit can be easily found and provide a check
of our results.
The large Higgs mass limit for the process $bg\rightarrow bH$ 
is
obtained by noting that the triangle and box diagrams shown in Figs. 
\ref{fg:self} -\ref{fg:box}
are of ${\cal O}\biggl({m_b^2\over v^2}\biggr)$
relative to the tree level amplitude.
 The only contributions which are enhanced by 
factors of ${M_H^2\over v^2}$
come from the renormalization of the $b\overline {b} H$ vertex,
Eq. \ref{yukdef}.  In the large $M_H$ limit,
\begin{equation}
L_{YUK}\rightarrow {m_b e\over 2 M_W}
\biggl({Z_H\over 1+\delta M_W^2/M_W^2}\biggr)
{\overline b} b H +{\rm terms~ not~ enhanced~ by~}{M_H^2\over v^2}
\, .
\end{equation}

The large $M_H$ limit of the Higgs wavefunction renormalization
is\cite{Marciano:1987un,Dawson:1989up},
\begin{equation}
Z_H=1+{1\over 16\pi^2}{M_H^2\over v^2}\biggl(6-\pi\sqrt{3}\biggr)
\, .
\end{equation}
Similarly,
the $W$ mass renormalization receives a contribution
proportional to $M_H^2$,
\begin{equation}
{\delta M_W^2\over M_W^2}=-{1\over 32\pi^2}{M_H^2\over v^2}\, .
\end{equation}
The leading $M_H$ corrections to $bH$ production are 
therefore\cite{Marciano:1987un},
\begin{eqnarray}
\sigma(bg\rightarrow bH)
&\rightarrow& {Z_H\over 1+{\delta M_W^2\over M_W^2}}
\sigma(bg\rightarrow bH)_0
\nonumber \\
&=&\biggl(1+{1\over 32\pi^2}
{M_H^2\over v^2}
\biggl[13-{2\pi \sqrt{3}}
\biggr]\biggr)
\sigma(bg\rightarrow bH)_0
\nonumber \\
&\equiv &\biggl(1+\Delta_{EW}^{M_H\rightarrow \infty}
\biggr)
\sigma(bg\rightarrow bH)_0\, .
\label{largemhdef}
\end{eqnarray}

\section{Results}
\label{results}
\subsection{Numerical Results}

\begin{figure}[t]
\begin{center}
\includegraphics[scale=0.6,angle=-90]{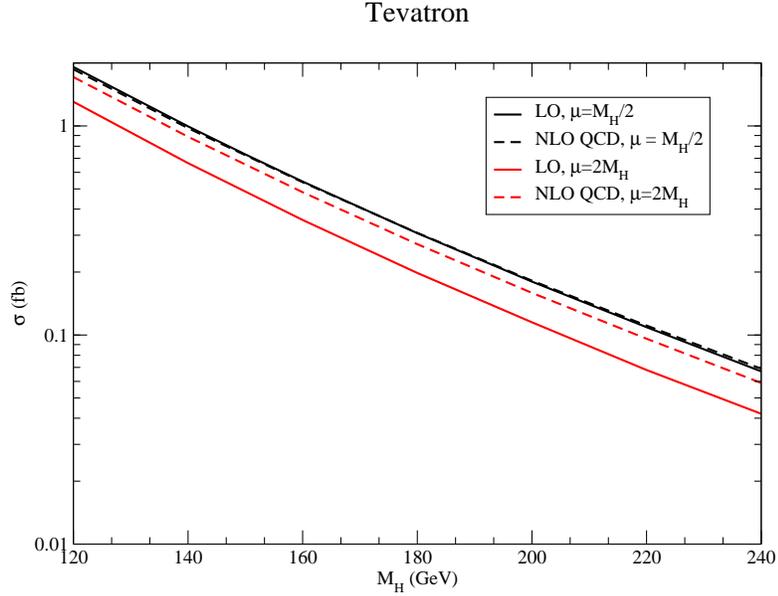}
\caption[]{Lowest order and NLO QCD 
results for $p {\overline p}\rightarrow b ({\overline b})
H X$ at the Tevatron
with $\sqrt{s}=1.96~TeV$  with $\sqrt{s}=1.96~GeV$, $p_{T}^b>20~GeV$, $\mid
\eta_b\mid < 2$, and $\Delta R > .4$. The renormalization/factorization
scales are set equal to $\mu$.}
\label{fg:tev2}
\end{center}
\end{figure}

\begin{figure}[t]
\begin{center}
\includegraphics[scale=0.6]{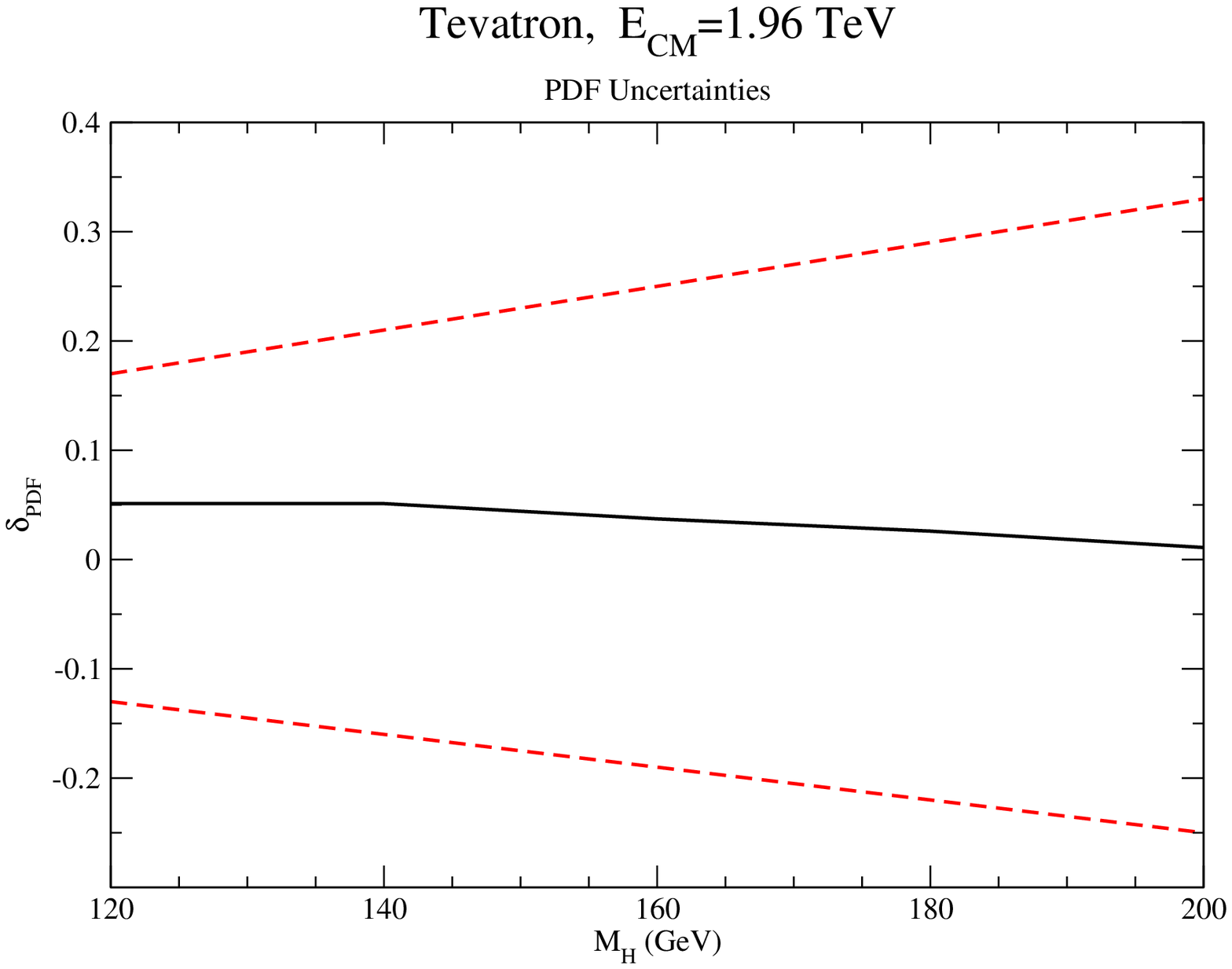}
\caption[]{PDF uncertainties
 for $p {\overline p}\rightarrow b ({\overline b})
H X$ at the Tevatron
with $\sqrt{s}=1.96~TeV$  $p_{T}^b>20~GeV$, $\mid
\eta_b\mid < 2$, $\Delta R > .4$, and $\mu=M_H/2$.
The solid line is 
$\sigma_{NLO}(CTEQ6.6)/\sigma_{NLO}(MRSW)-1$.
The dashed curves are the percentage variations from 
the central prediction between the upper and
lower predictions obtained using the CTEQ6.6 PDF error sets.}
\label{fg:tev3}
\end{center}
\end{figure}

\begin{figure}[t]
\begin{center}
\vskip 3in
\includegraphics[scale=0.6,angle=-90]{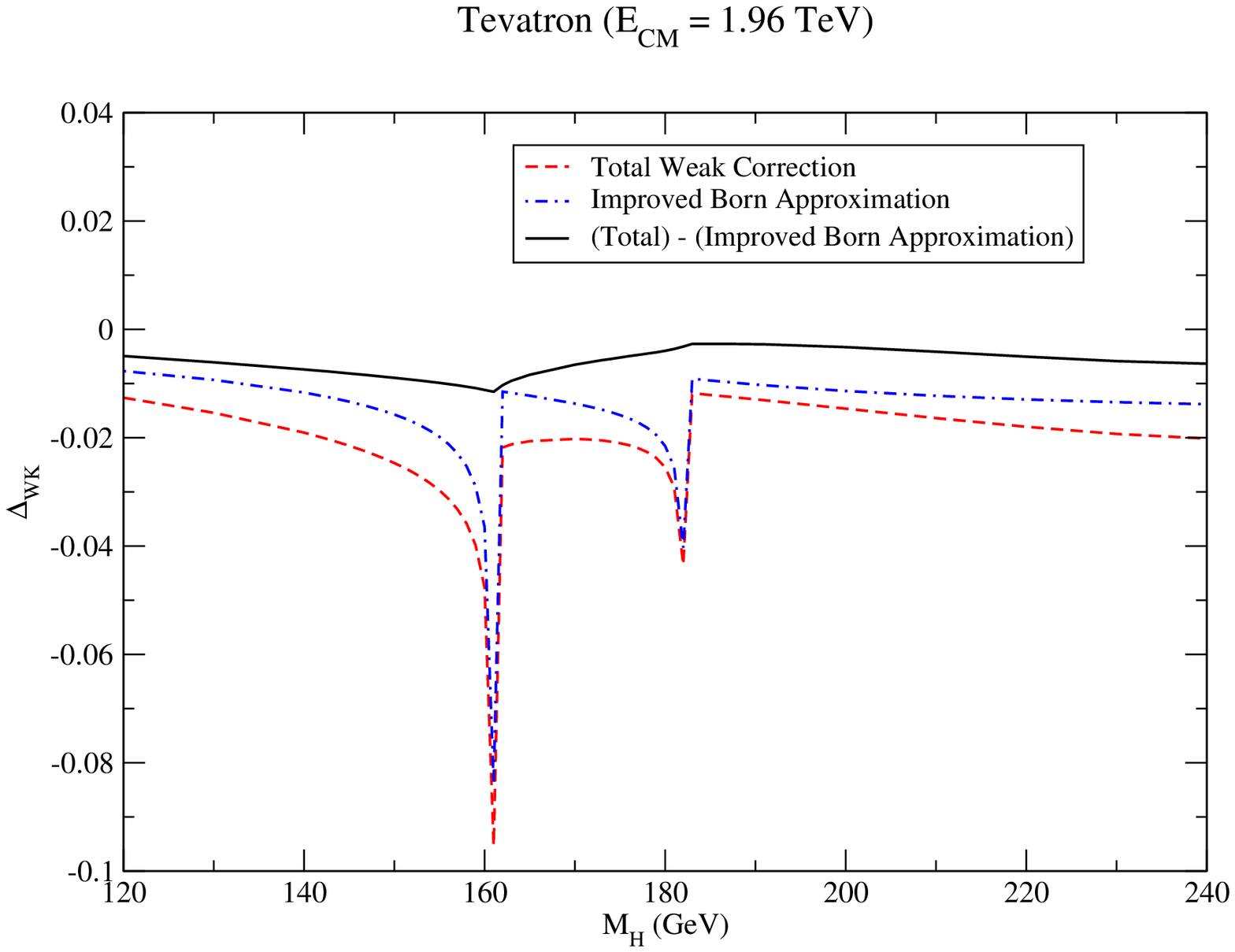}
\caption[]{
Tevatron results for the weak
corrections to  $p {\overline p}\rightarrow b ({\overline b})
H $ with $\sqrt{s}=1.96~GeV$, $p_{T}^b>20~GeV$, and $\mid
\eta_b\mid < 2$. The solid black curve represents the contributions which
cannot be factorized into an effective ${\overline b} b H$ vertex 
contribution and is always less than $1\%$.}
\label{fg:tev1}
\end{center}
\end{figure}

For our numerical studies, we use the following inputs:
\begin{eqnarray}
\alpha&=& 1/137.03599911\nonumber \\
G_\mu&=& 1.16637 \times 10^{-5}~GeV^{-2}\nonumber \\
M_Z&=&91.1875~GeV\nonumber \\
M_t&=&173.1~GeV\, .
\end{eqnarray}
We set the CKM mixing matrix to unity.
For the pole mass of the $b$ quark, we take $M_b=4.25~ GeV$.
 We use CTEQ6.6 PDFs\cite{Nadolsky:2008zw}
and vary the renormalization/factorization scales
 $\mu=\mu_R=\mu_F$ from $M_H/2$ to $2M_H$ in the total cross section
results.  

Our results are expressed as,
\begin{equation}
\sigma(bg\rightarrow bH)_{NLO}(\mu)=\sigma(bg\rightarrow bH)_0(\mu)\biggl(
1+\Delta_{QCD}(\mu)+\Delta_{QED}(\mu)+\Delta_{WK}(\mu)\biggr)\, ,
\label{cordef}
\end{equation}
where $\sigma_0$ is the Born cross section of Eq. \ref{eq:lo}
 evaluated with the
$1-$ loop renormalization group improved value for ${\overline{g_b}}(\mu)$
and includes the full $m_b$ mass dependence of Eq. \ref{eq:lo}. (Including
the $m_b$ dependence has almost no numerical effect).

The NLO QCD corrections are parameterized by the factor $\Delta_{QCD}$
and $\sigma_{NLO}$ is evaluated with the 
$2-$ loop renormalization group improved value for 
${\overline{g_b}}(\mu)$\footnote{Our results agree with those obtained
from MCFM\cite{mcfm}
and in Ref. \cite{Dawson:2005vi}
 and are presented here only to facilitate comparison with
$\Delta_{WK}$.}.  The NLO QCD corrections to the $bg\rightarrow bH$
process have been previously found in the S-ACOT scheme,
which
includes all effects of the finite $b$ mass to ${\cal O}(\alpha_s)$.
In the S-ACOT scheme\cite{Kramer:2000hn}, 
effects of a non-zero $b$ quark mass in
the process  $bg\rightarrow bH$ are absorbed into the definition of
the PDFs and to ${\cal O}(\alpha_s)$, we have schematically,
\begin{eqnarray}
\sigma_{NLO,QCD}(pp\rightarrow bH)_{m_b\ne 0}&\equiv&
\sigma_{LO}(pp\rightarrow bH)_{m_b=0}(1+\delta_{QCD})
+{\cal O}(\alpha_s^2)
\nonumber \\
\Delta_{QCD}&=&{\sigma_{LO}(pp\rightarrow bH)_{m_b=0}
\over\sigma_{LO}(pp\rightarrow bH)_{m_b\ne 0}}\delta_{QCD}\, ,
\end{eqnarray}
where $\delta_{QCD}$ is given in Ref. \cite{Campbell:2002zm}.
Both the CTEQ and MRSW PDF sets employ the S-ACOT scheme and so
our inclusion of $b$ mass effects is consistent to ${\cal O}(\alpha_s)$.

The QED and weak corrections are contained 
in $\Delta_{QED}$ and $\Delta_{WK}$, respectively.
As discussed above, we do not present results for $\Delta_{QED}$, but
assume they are negligible. The contribution of $\Delta_{WK}$ results
from the interference of the tree level amplitude with the $1$-loop
amplitudes shown above, which are generated numerically.
We compare our exact results of Eq. \ref{cordef}
with an ``Improved Born Approximation'', IBA, which is obtained by replacing
the tree level ${\overline b}bH$ vertex
of Eq. \ref{yukdef}  with the on-shell one loop electroweak corrected
vertex which can be found from the corrections to the decay $H\rightarrow
b{\overline b}$\cite{Kniehl:1991ze,Dabelstein:1991ky},
\begin{equation}
\Gamma(H\rightarrow b {\overline b})
=\Gamma(H\rightarrow b{\overline b})_0\biggl(1
+\Delta_{QCD}^{bbH}+\Delta_{QED}^{bbH}
+\Delta_{WK}^{bbH}\biggr)\, .
\end{equation}
We define the Improved Born Approximation in an obvious fashion  as
\begin{equation}
\sigma(bg\rightarrow bH)_{IBA}(\mu)\equiv
\sigma(bg\rightarrow bH)_0(\mu)\biggl(
1+\Delta_{QCD}^{bbH}
+\Delta_{EM}^{bbH}+\Delta_{WK}^{bbH}\biggr)\, .
\label{ibadef}
\end{equation}
The IBA approximation assumes that the bulk of the weak corrections
modify the ${\overline b} b H$ vertex.
In the case of the SUSY QCD corrections
 to $bg\rightarrow bH$ from squarks and gluinos, the
Improved Born Approximation is an excellent approximation to the full
rate\cite{Dawson:2007ur}. 

Results for the Tevatron are shown in Figs. \ref{fg:tev2}, \ref{fg:tev3},
and
\ref{fg:tev1}.
The Tevatron plots have $\sqrt{s}=1.96~TeV$, $\mid\eta_b\mid < 2.0$
and require $p_T^b>20~GeV$.  The NLO QCD corrections combine partons
if $\Delta R\equiv \sqrt{(\Delta \phi)^2+(\Delta\eta)^2} < 0.4$.
For $M_H=160$ GeV, the scale uncertainty at NLO with a variation
from $\mu=M_H/2$ to $2M_H$  is $\sim 10\%$, while for $M_H
=120~GeV$ it is $\sim 8\%$.  The PDF uncertainties are estimated
in Fig. \ref{fg:tev3} where we compare the CTEQ6.6 predictions with
those obtained using the MRSW2008 NLO PDFs\cite{Martin:2009iq},
(with $\mu=M_H/2$), and find agreement between the $2$ PDF
sets to within
better than $5\%$.  The PDF uncertainties using the CTEQ6 error sets
are
also shown in Fig. \ref{fg:tev3} and are quite large, varying between 
$15\%$ and $20\%$ for the masses considered 
here\footnote{The PDF uncertainties obtained from the 
$40$ CTEQ PDF
error sets were previously
obtained in Ref. \cite{Dawson:2005vi}.}.

Fig.~\ref{fg:tev1} shows the size of the weak corrections
as defined by Eq. \ref{cordef}. We note that the $\mu$ dependence
of $\Delta_{WK}$ is extremely small.
 The weak corrections are well approximated by
the IBA of Eq. \ref{ibadef} (the dot-
dashed line of Fig. \ref{fg:tev1}), 
with the remaining corrections (the solid line in
Fig. \ref{fg:tev1} ) always less than $1\%$.  Except near the $W^+W$ and $ZZ$
resonances, 
$\Delta_{WK}$ in the Standard Model is
significantly smaller than the uncertainties from the QCD scale variation
and the PDF uncertainties.

\begin{figure}[t]
\begin{center}
\vskip 3in
\includegraphics[scale=0.6,angle=-90]{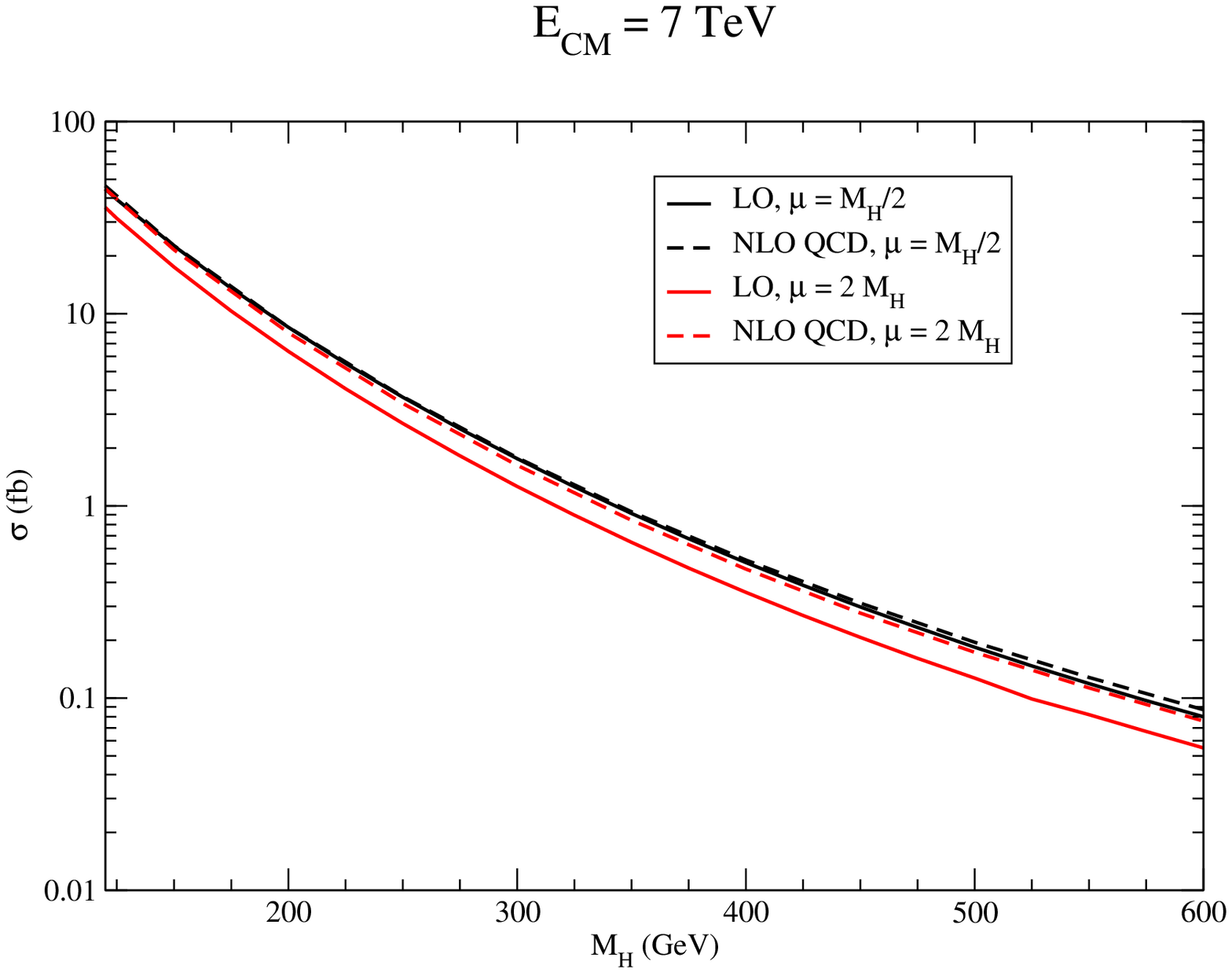}
\caption[]{
Lowest order and NLO QCD 
results for  $p p \rightarrow b ({\overline b})
H X$ at the 
LHC  with $\sqrt{s}=7~TeV$, $p_{T}^b>25~GeV$,  $\mid
\eta_b\mid < 2.5$, and $\Delta r > .4$.
The renormalization/factorization
scales are set equal to $\mu$.}
\label{fg:tev7}
\end{center}
\end{figure}

\begin{figure}[t]
\begin{center}
\vskip 3.in
\includegraphics[scale=0.6,angle=-90]{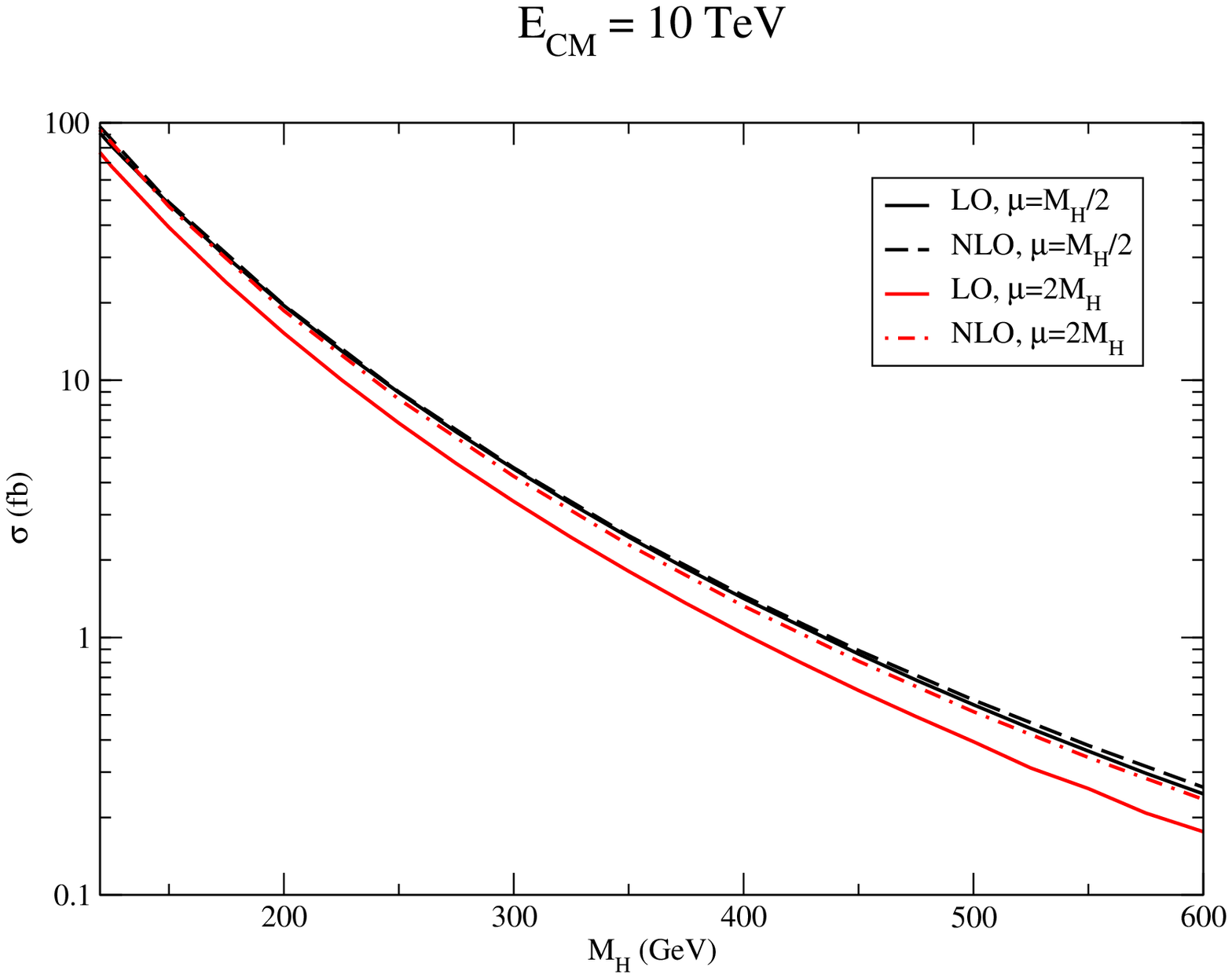}
\caption[]{
Lowest order and NLO QCD 
results for  $p p \rightarrow b ({\overline b})
H X$ at the 
LHC  with $\sqrt{s}=10~TeV$, $p_{T}^b>25~GeV$,  $\mid
\eta_b\mid < 2.5$, and $\Delta r > .4$.
The renormalization/factorization
scales are set equal to $\mu$.}
\label{fg:tev10}
\end{center}
\end{figure}

\begin{figure}[t]
\begin{center}
\includegraphics[scale=0.6]{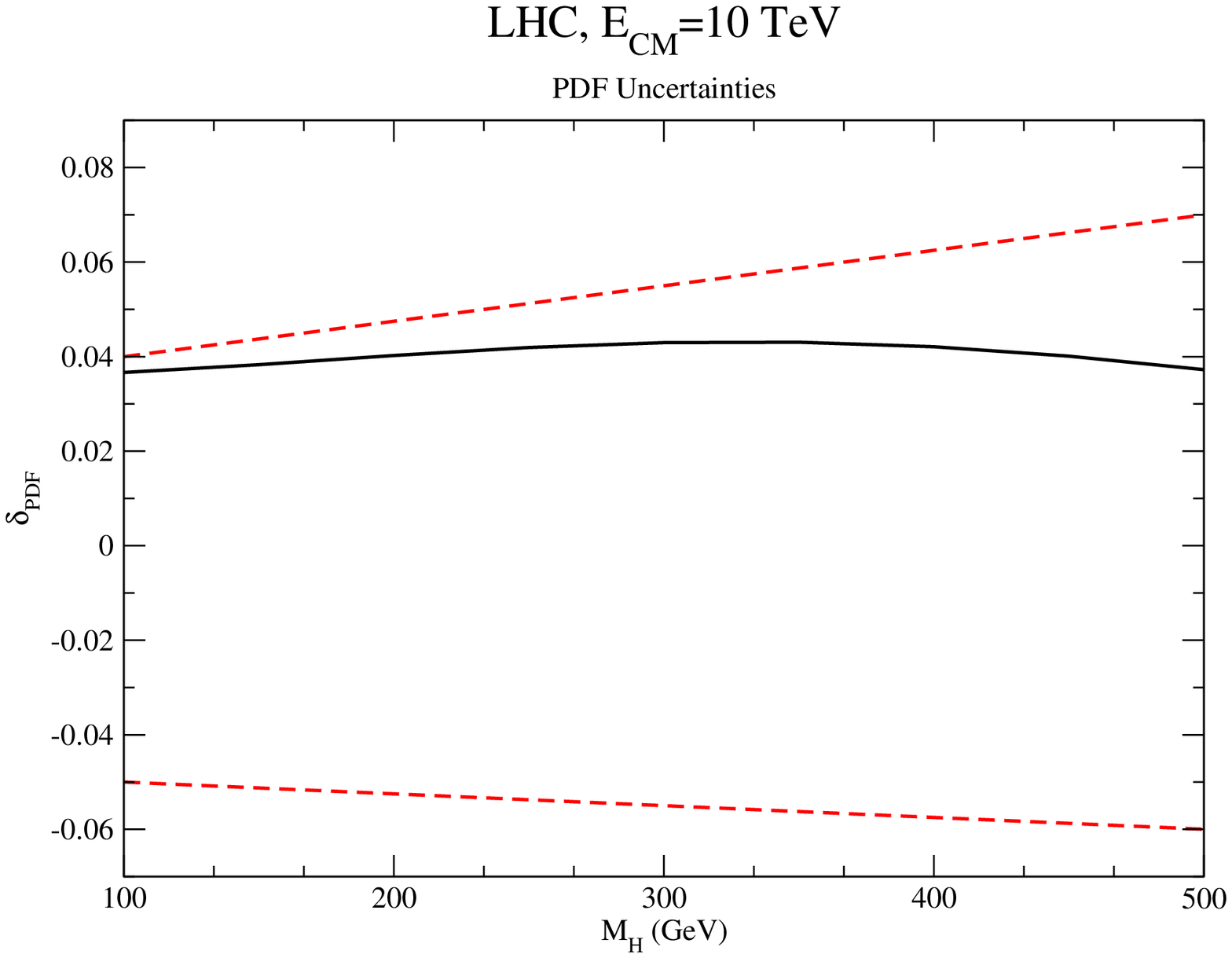}
\caption[]{PDF uncertainties
 for $p p \rightarrow b ({\overline b})
H $ at the LHC
with $\sqrt{s}=10~TeV$, $p_{T}^b>25~GeV$, $\mid
\eta_b\mid < 2.5$, $\Delta R > .4$, and $\mu=M_H/2$.
The solid line is 
$\sigma_{NLO}(CTEQ6.6)/\sigma_{NLO}(MRSW)-1$.
The dashed curves are the percentage variations from 
the central prediction between the upper and
lower predictions obtained using the CTEQ6.6 PDF error sets.}
\label{fg:tev10pdf}
\end{center}
\end{figure}

At the LHC, we consider 
$\sqrt{s}=7 ~TeV$ and $\sqrt{s}=10~TeV$, with  $\mid\eta_b\mid < 2.5$,
$p_{T}^b>25~GeV$ and 
$\Delta R>  0.4$.
The NLO QCD corrected cross
sections are shown in Figs. \ref{fg:tev7} and \ref{fg:tev10}.  The NLO cross
section is reduced by a factor of $\sim 2.2$  for $M_H=150~GeV$ (with
$\mu=M_H/2$) when going
from $\sqrt{s}=10~TeV$ to $7~TeV$.
At $\sqrt{s}=7~TeV$, and  $M_H=150$ GeV, 
the scale uncertainty at NLO with a variation
from $\mu=M_H/2$ to $2M_H$  is $\sim 5\%$, while for $M_H
=300~GeV$ it is $\sim 9\%$.
The PDF uncertainties for $\sqrt{s}=10~TeV$ are estimated
in Fig. \ref{fg:tev10pdf} where we compare the CTEQ6.6 predictions with
those obtained using the MRSW2008 NLO PDFs\cite{Martin:2009iq},
(with $\mu=M_H/2$), and find agreement between the $2$ PDF
sets to within
better than $3\%$.  The PDF uncertainties using the CTEQ6 error sets
are
also shown in Fig. \ref{fg:tev10pdf} and are
smaller than at the Tevatron, varying between 
$4\%$ and $6\%$ for the masses considered 
here.  The PDF uncertainties are similar for $\sqrt{s}=7~TeV$.

\begin{figure}[t]
\begin{center}
\includegraphics[scale=0.6]{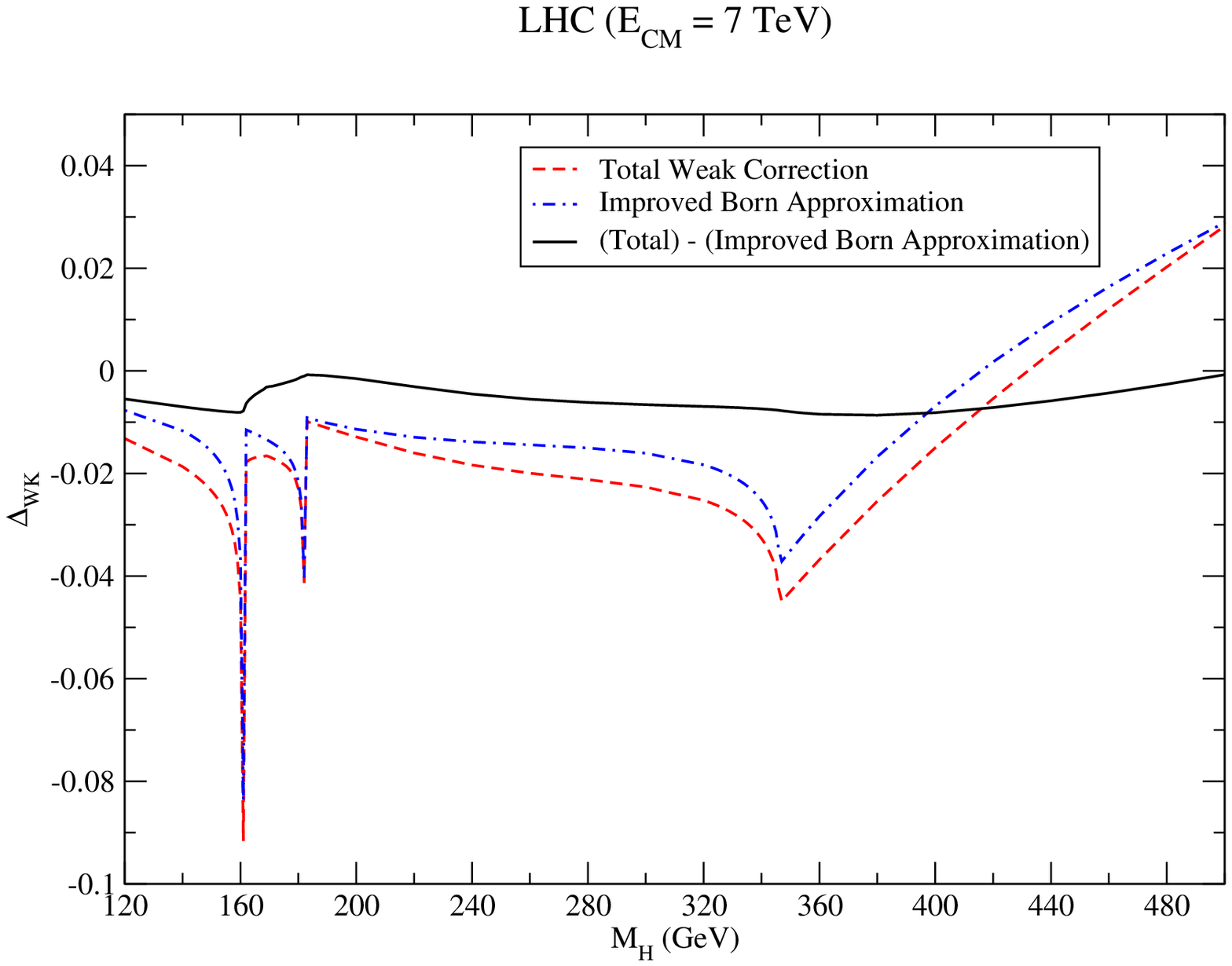}
\caption[]{
LHC results for the weak
corrections to  $p p \rightarrow b ({\overline b})
H $ with $\sqrt{s}=7~TeV$, $p_{T}^b>25~GeV$, and $\mid
\eta_b\mid < 2.5$. The solid black curve represents the contributions which
cannot be factorized into an effective ${\overline b} b H$ vertex contribution and is  less than $1\%$ for $M_H < 500~GeV$.}
\label{fg:tev7_ew}
\end{center}
\end{figure}

\begin{figure}[t]
\begin{center}
\includegraphics[scale=0.6]{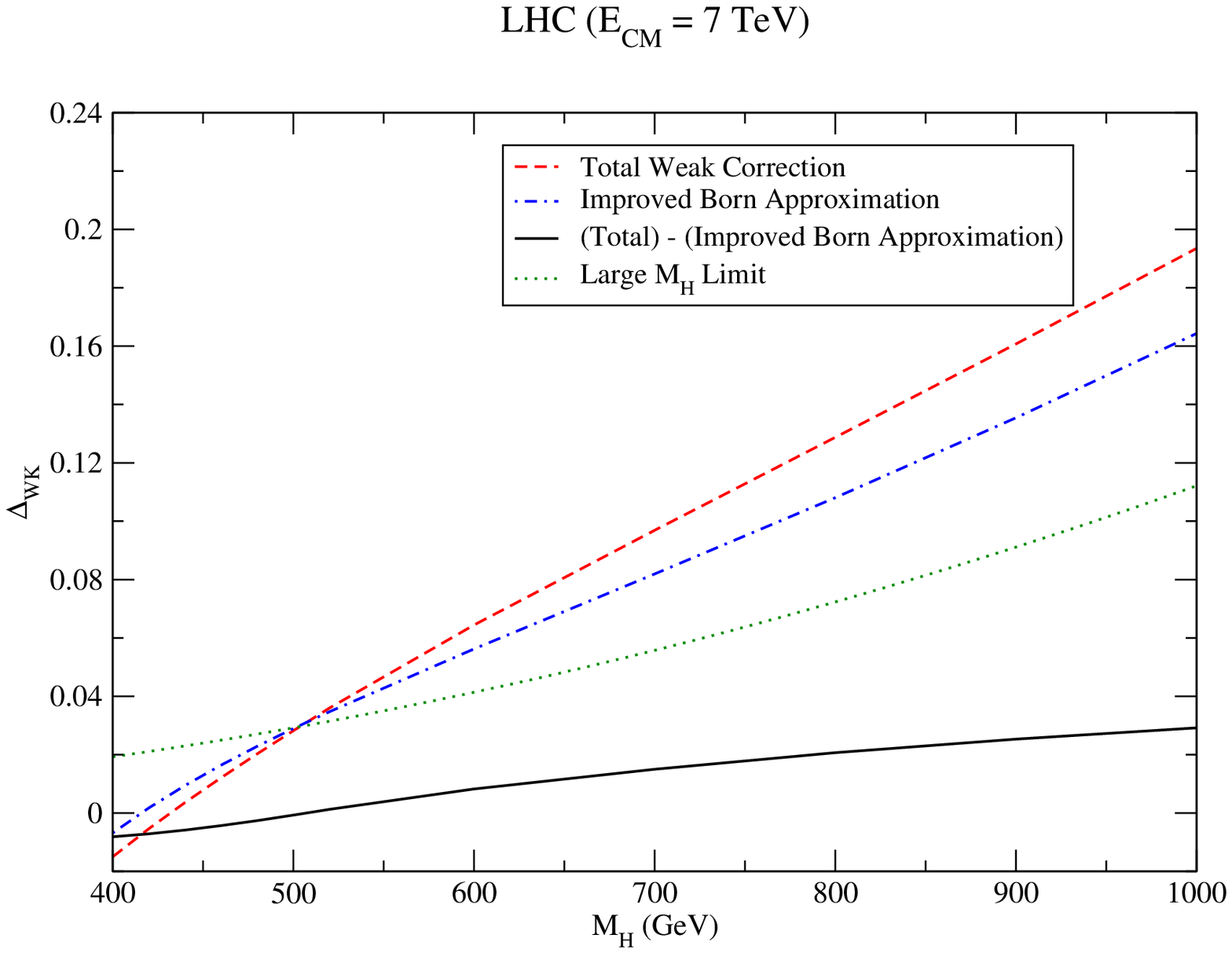}
\caption[]{
LHC results for the weak
corrections to  $p p \rightarrow b ({\overline b})
H $ with $\sqrt{s}=7~TeV$, $p_{T}^b>25~GeV$, and $\mid
\eta_b\mid < 2.5$. The solid black curve represents the contributions which
cannot be factorized into an effective ${\overline b} b H$ vertex 
contribution. The dotted line is the large Higgs mass limit of Eq.
 \ref{largemhdef}.}
\label{fg:tev72ew}
\end{center}
\end{figure}

\begin{figure}[t]
\begin{center}
\includegraphics[scale=0.6]{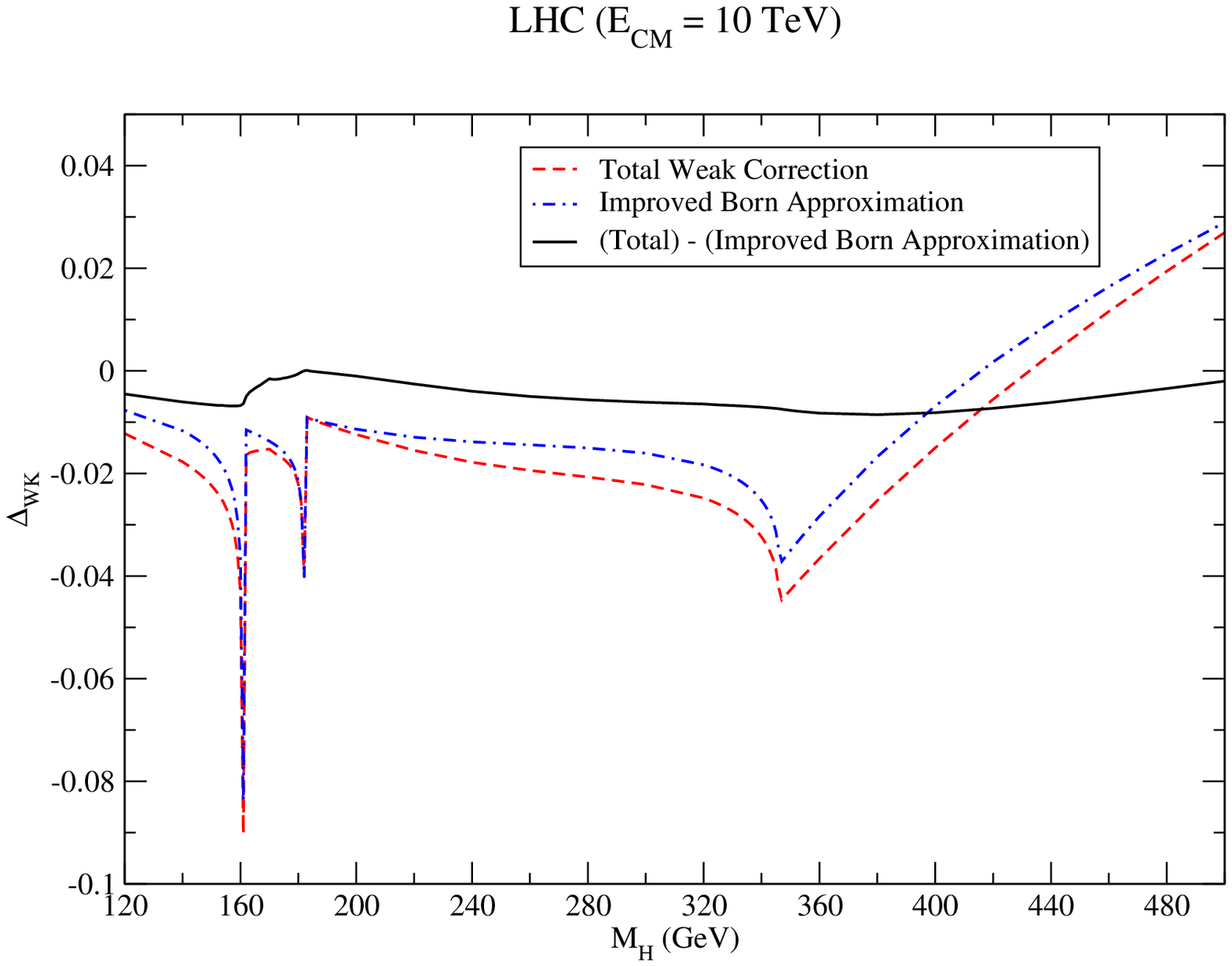}
\caption[]{
LHC results for the weak
corrections to  $p p \rightarrow b ({\overline b})
H $ with $\sqrt{s}=10~TeV$, $p_{T}^b>25~GeV$, and $\mid
\eta_b\mid < 2.5$. The solid black curve represents the contributions which
cannot be factorized into an effective ${\overline b} b H$ vertex 
contribution and is  less than $1\%$ for $M_H < 500~GeV$.}
\label{fg:tev10_ew}
\end{center}
\end{figure}

\begin{figure}[t]
\begin{center}
\includegraphics[scale=0.6]{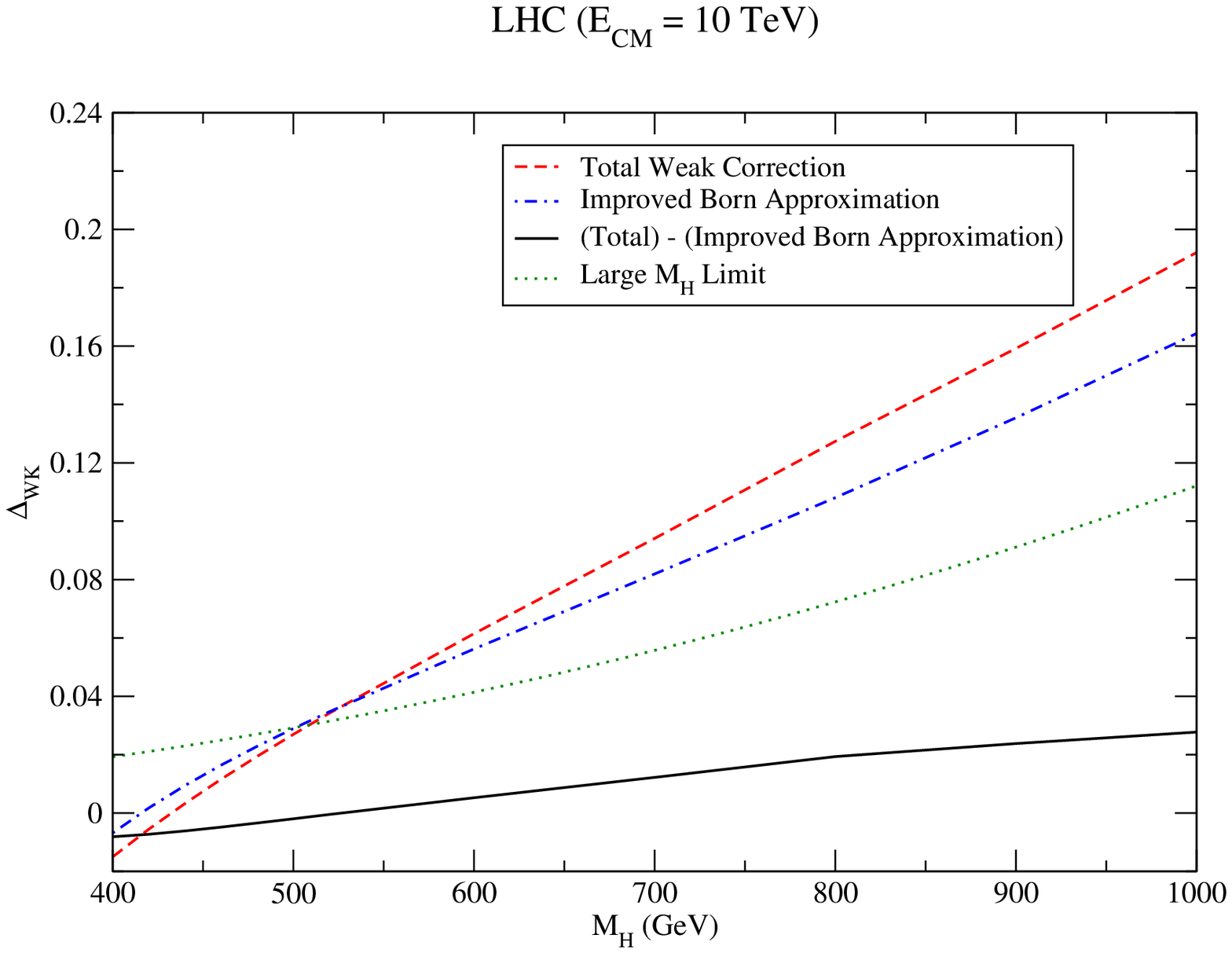}
\caption[]{
LHC results for the weak
corrections to  $p p \rightarrow b ({\overline b})
H $ with $\sqrt{s}=10~TeV$, $p_{T}^b>25~GeV$, and $\mid
\eta_b\mid < 2.5$. The solid black curve represents the contributions which
cannot be factorized into an effective ${\overline b} b H$ vertex 
contribution. The dotted line is the large Higgs mass limit of Eq.
 \ref{largemhdef}.}
\label{fg:tev102ew}
\end{center}
\end{figure}

 The weak corrections are shown in Figs.
\ref{fg:tev7_ew} and \ref{fg:tev10_ew} for $M_H < 500~GeV$.  The IBA 
(Eq. \ref{ibadef}) encapsulates the total weak
corrections to better than $1\%$ for $M_H < 500~GeV$.
We  show the weak effects for $M_H> 500~GeV$, along
with the large $M_H$  limit of Eq. 
\ref{largemhdef}, in Figs. \ref{fg:tev72ew} and 
\ref{fg:tev102ew}.  
For $M_H=1~TeV$, the IBA underestimates the total weak corrections by
about $3\%$ at $\sqrt{s}=10~TeV$.
For large $M_H$ ($M_H > 2M_t$), the weak corrections are
significant and are greater than $18\%$ for $M_H\sim 1~TeV$.
We note that the large $M_H$ limit underestimates the weak corrections 
by about $5\%$ at $M_H=1~TeV$, implying that the 
 $\log(M_H)$ terms are numerically important.  For heavy Higgs bosons,
$M_H >500~GeV$, the weak corrections are larger than uncertainties
from PDFs and the scale choice, and it is meanful to include them
in precision calculations.

\subsection{The $m_b=0$ Limit}
It is interesting to consider the $m_b\rightarrow 0$ limit of the
$bg\rightarrow bH$ amplitude.  In this limit, the $b$-Higgs Yukawa coupling
vanishes, $g_b=m_b/v\rightarrow 0$, and the tree level amplitude shown
in Fig. \ref{fg:lo} is identically zero.  The first non-zero contributions
to $bg\rightarrow bH$ with $m_b=0$
 arise from the squares of a subset of the $1-$ loop amplitudes
shown in Figs. \ref{fg:vert} and \ref{fg:box} 
and are ${\cal O}(\alpha_s G_F^3)$.  The contributions which are
non-zero in the $m_b\rightarrow 0$ limit involve the coupling of the
Higgs to either a top quark or a pair of gauge bosons (and the
corresponding Goldstone bosons).  
These contributions have been calculated in Ref. \cite{Mrenna:1995cf} and  
we have checked that the squares of our $1-$ loop amplitudes reproduce
their results in the $m_b=0$ limit.  
Since these diagrams are not suppressed by a small
$b$ quark Yukawa coupling, they give a comparatively large contribution.
At $\sqrt{s}=7~TeV$ 
and $M_H=120~GeV$, we find 
that the
${\cal O}(\alpha_s G_F^3)$ contribution with $m_b=0$ is around $8\%$ 
of the Born cross section 
shown in Fig.\ref{fg:tev7} with our cuts.

Although our calculations are purely
Standard Model, we are, however, motivated by a very different scenario than 
the authors of Ref. \cite{Mrenna:1995cf}. In models 
with an enhanced coupling of the $b$ quark
to a Higgs boson, the tree level amplitude can be significantly larger
than in the Standard Model.  In such models, it is important to
understand the numerical effect of the interference of the tree level
amplitude with the one-loop weak corrections.  Future work will explore
the role of the electroweak corrections in models with non-standard
$b$ quark Higgs Yukawa couplings, in particular the MSSM with large
$\tan\beta$.

\section{Conclusion}

We have computed the Standard
Model weak corrections to the processes $pp\rightarrow b (
{\overline b}) H$ at the LHC and 
$p{\overline p}\rightarrow b (
{\overline b}) H$ at the Tevatron.  In both cases, the results are well
approximated by including only 
 the on-shell ${\overline b} b H$ vertex corrections, with
the remaining weak corrections less than $1-2\%$ for $M_H < 500~GeV$.
This observation makes it straightforward to estimate the weak
effects
of non-Standard Model $b$ quark Yukawa couplings on the $bH$ production
process.

At the Tevatron, the weak effects are always much smaller than
scale and PDF uncertainties and so can be neglected in the Standard
Model.
At the LHC, for large $M_H$ the weak corrections can
become significant and can be larger than scale and PDF uncertainties.  
At the LHC with $\sqrt{s}=10~TeV$, the
corrections of ${\cal O}\biggl({M_H^2\over v^2}\biggr)$ are $\sim 18\%$
for $M_H=1~TeV$.

\section*{Acknowledgements}
We thank Chris Jackson, Laura Reina, Christian Sturm and Doreen
Wackeroth for many helpful discussions.
This work is supported by the United States Department of Energy under
Grant DE-AC02-98CH10886.

\bibliography{bg_ew}

\begin{thebibliography}{10}

\bibitem{Dawson:2005vi}
S.~Dawson, C.~B. Jackson, L.~Reina, and D.~Wackeroth.
\newblock Higgs production in association with bottom quarks at hadron
  colliders.
\newblock {\em Mod. Phys. Lett.}, A21:89--110, 2006.

\bibitem{Dawson:2004sh}
S.~Dawson, C.~B. Jackson, L.~Reina, and D.~Wackeroth.
\newblock Higgs boson production with one bottom quark jet at hadron colliders.
\newblock {\em Phys. Rev. Lett.}, 94:031802, 2005.

\bibitem{Campbell:2004pu}
J.~Campbell et~al.
\newblock Higgs boson production in association with bottom quarks.
\newblock 2004.

\bibitem{Dittmaier:2003ej}
Stefan Dittmaier, Michael Kramer, and Michael Spira.
\newblock Higgs radiation off bottom quarks at the tevatron and the lhc.
\newblock {\em Phys. Rev.}, D70:074010, 2004.

\bibitem{Dawson:2003kb}
S.~Dawson, C.~B. Jackson, L.~Reina, and D.~Wackeroth.
\newblock Exclusive higgs boson production with bottom quarks at hadron
  colliders.
\newblock {\em Phys. Rev.}, D69:074027, 2004.

\bibitem{Campbell:2002zm}
John~M. Campbell, R.~Keith Ellis, F.~Maltoni, and S.~Willenbrock.
\newblock {Higgs boson production in association with a single bottom quark}.
\newblock {\em Phys. Rev.}, D67:095002, 2003.

\bibitem{Maltoni:2005wd}
Fabio Maltoni, Thomas McElmurry, and Scott Willenbrock.
\newblock Inclusive production of a higgs or z boson in association with heavy
  quarks.
\newblock {\em Phys. Rev.}, D72:074024, 2005.

\bibitem{Dicus:1998hs}
D.~Dicus, T.~Stelzer, Z.~Sullivan, and S.~Willenbrock.
\newblock Higgs boson production in association with bottom quarks at
  next-to-leading order.
\newblock {\em Phys. Rev.}, D59:094016, 1999.

\bibitem{Maltoni:2003pn}
F.~Maltoni, Z.~Sullivan, and S.~Willenbrock.
\newblock Higgs-boson production via bottom-quark fusion.
\newblock {\em Phys. Rev.}, D67:093005, 2003.

\bibitem{Brein:2003df}
Oliver Brein and Wolfgang Hollik.
\newblock Mssm higgs bosons associated with high-p(t) jets at hadron colliders.
\newblock {\em Phys. Rev.}, D68:095006, 2003.

\bibitem{Field:2003yy}
B.~Field, S.~Dawson, and J.~Smith.
\newblock {Scalar and pseudoscalar Higgs boson plus one jet production at the
  LHC and Tevatron}.
\newblock {\em Phys. Rev.}, D69:074013, 2004.

\bibitem{Carena:1998gk}
Marcela~S. Carena, S.~Mrenna, and C.~E.~M. Wagner.
\newblock Mssm higgs boson phenomenology at the tevatron collider.
\newblock {\em Phys. Rev.}, D60:075010, 1999.

\bibitem{Carena:2007aq}
Marcela~S. Carena, A.~Menon, and C.~E.~M. Wagner.
\newblock Challenges for mssm higgs searches at hadron colliders.
\newblock {\em Phys. Rev.}, D76:035004, 2007.

\bibitem{Barnett:1987jw}
R.~Michael Barnett, Howard~E. Haber, and Davison~E. Soper.
\newblock Ultraheavy particle production from heavy partons at hadron
  colliders.
\newblock {\em Nucl. Phys.}, B306:697, 1988.

\bibitem{Olness:1987ep}
Fredrick~I. Olness and Wu-Ki Tung.
\newblock When is a heavy quark not a parton? charged higgs production and
  heavy quark mass effects in the qcd based parton model.
\newblock {\em Nucl. Phys.}, B308:813, 1988.

\bibitem{Harlander:2003ai}
Robert~V. Harlander and William~B. Kilgore.
\newblock Higgs boson production in bottom quark fusion at next-to-
  next-to-leading order.
\newblock {\em Phys. Rev.}, D68:013001, 2003.

\bibitem{Dittmaier:2006cz}
Stefan Dittmaier, Michael Kramer, 1, Alexander Muck, and Tobias Schluter.
\newblock {MSSM Higgs-boson production in bottom-quark fusion: Electroweak
  radiative corrections}.
\newblock {\em JHEP}, 03:114, 2007.

\bibitem{Hollik:2006vn}
Wolfgang Hollik and Michael Rauch.
\newblock {Higgs-Boson Production in Association with Heavy Quarks}.
\newblock {\em AIP Conf. Proc.}, 903:117--120, 2007.

\bibitem{Dawson:2007ur}
S.~Dawson and C.~B. Jackson.
\newblock {SUSY QCD Corrections to Associated Higgs-bottom Quark Production}.
\newblock {\em Phys. Rev.}, D77:015019, 2008.

\bibitem{Abazov:2008hh}
V.~M. Abazov et~al.
\newblock {Search for neutral Higgs bosons in multi-b-jet events in $p \bar{p}$
  collisions at $\sqrt{s}$ = 1.96-TeV}.
\newblock {\em Phys. Rev. Lett.}, 101:221802, 2008.

\bibitem{Abazov:2008zz}
V.~M. Abazov et~al.
\newblock {Search for neutral Higgs bosons $tan\beta$ in the b(h/H/A) $\to b
  \tau \tau$ channel }.
\newblock {\em Phys. Rev. Lett.}, 102:051804, 2009.

\bibitem{Boudjema:2008zn}
Fawzi Boudjema and Le~Duc Ninh.
\newblock {b anti-b Higgs production at the LHC: Yukawa corrections and the
  leading Landau singularity}.
\newblock {\em Phys. Rev.}, D78:093005, 2008.

\bibitem{Mrenna:1995cf}
S.~Mrenna and C.~P. Yuan.
\newblock {High $p_{T}$ Higgs boson production at hadron colliders to O
  (alpha-s G(F) (3) )}.
\newblock {\em Phys. Rev.}, D53:3547--3554, 1996.

\bibitem{Kniehl:1994ju}
Bernd~A. Kniehl and Michael Spira.
\newblock {Two loop O (alpha-s G(F) $m($t$) ^{2}$ correction to the $H \to b
  \bar{b}$ decay rate}.
\newblock {\em Nucl. Phys.}, B432:39--48, 1994.

\bibitem{Djouadi:2005gi}
Abdelhak Djouadi.
\newblock {The Anatomy of electro-weak symmetry breaking. I: The Higgs boson in
  the standard model}.
\newblock {\em Phys. Rept.}, 457:1--216, 2008.

\bibitem{Hollik:1988ii}
W.~F.~L. Hollik.
\newblock {Radiative Corrections in the Standard Model and their Role for
  Precision Tests of the Electroweak Theory}.
\newblock {\em Fortschr. Phys.}, 38:165--260, 1990.

\bibitem{Bardin:1999ak}
Dmitri~Yu. Bardin and G.~Passarino.
\newblock {The standard model in the making: Precision study of the electroweak
  interactions}.
\newblock Oxford, UK: Clarendon (1999) 685 p.

\bibitem{Chen:2008jg}
Mu-Chun Chen, Sally Dawson, and C.~B. Jackson.
\newblock {Higgs Triplets, Decoupling, and Precision Measurements}.
\newblock {\em Phys. Rev.}, D78:093001, 2008.

\bibitem{Chen:2003fm}
Mu-Chun Chen and Sally Dawson.
\newblock {One-loop radiative corrections to the rho parameter in the littlest
  Higgs model}.
\newblock {\em Phys. Rev.}, D70:015003, 2004.

\bibitem{Burkhardt:2001xp}
H.~Burkhardt and B.~Pietrzyk.
\newblock {Update of the hadronic contribution to the QED vacuum polarization}.
\newblock {\em Phys. Lett.}, B513:46--52, 2001.

\bibitem{Sirlin:1981yz}
A.~Sirlin and W.~J. Marciano.
\newblock {Radiative Corrections to Muon-neutrino N $\to$ mu- X and their
  Effect on the Determination of rho**2 and sin**2- Theta(W)}.
\newblock {\em Nucl. Phys.}, B189:442, 1981.

\bibitem{Sirlin:1980nh}
A.~Sirlin.
\newblock {Radiative Corrections in the SU(2)-L x U(1) Theory: A Simple
  Renormalization Framework}.
\newblock {\em Phys. Rev.}, D22:971--981, 1980.

\bibitem{Marciano:1983wwa}
William~J. Marciano and A.~Sirlin.
\newblock {Testing the Standard Model by Precise Determinations of W+- and Z
  Masses}.
\newblock {\em Phys. Rev.}, D29:945, 1984.

\bibitem{Kniehl:1991ze}
Bernd~A. Kniehl.
\newblock {Radiative corrections for $H \to$ f anti-f ($\gamma$) in the
  standard model}.
\newblock {\em Nucl. Phys.}, B376:3--28, 1992.

\bibitem{Larin:1995sq}
S.~A. Larin, T.~van Ritbergen, and J.~A.~M. Vermaseren.
\newblock {The Large top quark mass expansion for Higgs boson decays into
  bottom quarks and into gluons}.
\newblock {\em Phys. Lett.}, B362:134--140, 1995.

\bibitem{Chetyrkin:1995pd}
K.~G. Chetyrkin and A.~Kwiatkowski.
\newblock {Second order QCD corrections to scalar and pseudoscalar Higgs decays
  into massive bottom quarks}.
\newblock {\em Nucl. Phys.}, B461:3--18, 1996.

\bibitem{Drees:1990dq}
Manuel Drees and Ken-ichi Hikasa.
\newblock {NOTE ON QCD CORRECTIONS TO HADRONIC HIGGS DECAY}.
\newblock {\em Phys. Lett.}, B240:455, 1990.

\bibitem{Braaten:1980yq}
E.~Braaten and J.~P. Leveille.
\newblock {Higgs Boson Decay and the Running Mass}.
\newblock {\em Phys. Rev.}, D22:715, 1980.

\bibitem{Dabelstein:1991ky}
A.~Dabelstein and W.~Hollik.
\newblock {Electroweak corrections to the fermionic decay width of the standard
  Higgs boson}.
\newblock {\em Z. Phys.}, C53:507--516, 1992.

\bibitem{Denner:2005fg}
Ansgar Denner, S.~Dittmaier, M.~Roth, and L.~H. Wieders.
\newblock {Electroweak corrections to charged-current e+ e- --> 4 fermion
  processes: Technical details and further results}.
\newblock {\em Nucl. Phys.}, B724:247--294, 2005.

\bibitem{Passarino:2010qk}
Giampiero Passarino, Christian Sturm, and Sandro Uccirati.
\newblock {Higgs Pseudo-Observables, Second Riemann Sheet and All That}.
\newblock 2010.

\bibitem{Martin:2004dh}
A.~D. Martin, R.~G. Roberts, W.~J. Stirling, and R.~S. Thorne.
\newblock {Parton distributions incorporating QED contributions}.
\newblock {\em Eur. Phys. J.}, C39:155--161, 2005.

\bibitem{Hahn:2000kx}
Thomas Hahn.
\newblock {Generating Feynman diagrams and amplitudes with FeynArts 3}.
\newblock {\em Comput. Phys. Commun.}, 140:418--431, 2001.

\bibitem{Hahn:1998yk}
T.~Hahn and M.~Perez-Victoria.
\newblock {Automatized one-loop calculations in four and D dimensions}.
\newblock {\em Comput. Phys. Commun.}, 118:153--165, 1999.

\bibitem{Marciano:1987un}
William~J. Marciano and Scott S.~D. Willenbrock.
\newblock {RADIATIVE CORRECTIONS TO HEAVY HIGGS SCALAR PRODUCTION AND DECAY}.
\newblock {\em Phys. Rev.}, D37:2509, 1988.

\bibitem{Dawson:1989up}
Sally Dawson and Scott Willenbrock.
\newblock {RADIATIVE CORRECTIONS TO LONGITUDINAL VECTOR BOSON SCATTERING}.
\newblock {\em Phys. Rev.}, D40:2880, 1989.

\bibitem{Nadolsky:2008zw}
Pavel~M. Nadolsky et~al.
\newblock {Implications of CTEQ global analysis for collider observables}.
\newblock {\em Phys. Rev.}, D78:013004, 2008.

\bibitem{mcfm}
http://mcfm.fnal.gov.

\bibitem{Kramer:2000hn}
Michael Kramer, 1, Fredrick~I. Olness, and Davison~E. Soper.
\newblock {Treatment of heavy quarks in deeply inelastic scattering}.
\newblock {\em Phys. Rev.}, D62:096007, 2000.

\bibitem{Martin:2009iq}
A.~D. Martin, W.~J. Stirling, R.~S. Thorne, and G.~Watt.
\newblock {Parton distributions for the LHC}.
\newblock {\em Eur. Phys. J.}, C63:189--285, 2009.

\end{thebibliography}
\bibliographystyle{unsrt}

\end{document}